\documentclass[twocolumn,showpacs,preprintnumbers,nofootinbib,amsmath,amssymb]{revtex4}
\usepackage{hyperref,slashed,color}
\usepackage{graphicx,subfig}

\begin{document}
\preprint{TUHEP-TH-08165}
\title{Electroweak Chiral Lagrangian from Natural Topcolor-assisted Technicolor Model}

\bigskip
\author{Jun-Yi
Lang$^{1,2}$\footnote{Email:\href{mailto:lang00@mails.tsinghua.edu.cn}{lang00@mails.tsinghua.edu.cn}.},
Shao-Zhou Jiang$^{1,2}$\footnote{Email:
\href{mailto:jsz@mails.tsinghua.edu.cn}{jsz@mails.tsinghua.edu.cn}.},
 and Qing Wang$^{1,2}$\footnote{Email:
\href{mailto:wangq@mail.tsinghua.edu.cn}{wangq@mail.tsinghua.edu.cn}.}\footnote{corresponding
author}\\~}

\bigskip
\affiliation{$^1$Center for High Energy Physics, Tsinghua University, Beijing 100084, China\\
$^2$Department of Physics, Tsinghua University, Beijing 100084,
China\footnote{mailing address}}

\begin{abstract}
Based on previous studies computing coefficients of the electroweak
chiral Lagrangian from C.T.Hill's schematic topcolor-assisted
technicolor model, we generalize the calculation to K.Lane's
prototype natural topcolor-assisted technicolor model.  We find that
typical features of the model are qualitatively similar as those of
Hill's model, but Lane's model prefers smaller technicolor group and
$Z'$ mass must be smaller than $400$GeV, further $S$ parameter is
around order of $+1$ mainly due to existence of three doublets of
techniquarks. We obtain the values for all coefficients of the
electroweak chiral Lagrangian up to order of $p^4$. Apart from
 negative large four fermion coupling values, ETC impacts on the electroweak
chiral Lagrangian coefficients are small, since techniquark self
energy which determines these coefficients in general receives
almost no influence from ETC induced four fermion interactions
except for its large momentum tail.
\end{abstract}
 \maketitle
\section{Introduction}

Topcolor assisted technicolor (TC2) model realizes the electroweak
symmetry breaking (EWSB) by joining technicolor (TC) and topcolor
together to remove the objections that topcolor is unnatural and
that TC cannot generate a large top mass. In the first schematic
model proposed by C.T.Hill\cite{Hill95}, EWSB is driven mainly by TC
interactions and light quark and lepton masses are expected to be
generated by extended technicolor (ETC). The third generation
$(t,b)_{L,R}$ is arranged to transform with the usual quantum
numbers under the gauge group $SU(3)_1\otimes U(1)_1$ while
$(u,d),(c,s)$ transform under a separate group $SU(3)_2\otimes
U(1)_2$. At a scale of order 1TeV, $SU(3)_1\otimes SU(3)_2\otimes
U(1)_1\otimes U(1)_2$ is dynamically broken to the diagonal subgroup
$SU(3)_C\otimes U(1)_Y$,
 and $SU(3)_1\otimes U(1)_1$ interactions are supercritical for $t$
quark leading top condensation, but subcritical for $b$ quark
causing no bottom condensation which achieve large mass difference
between $t$ and $b$ quarks.

 As a candidate of new physics model, before any new particles such as
 $Z'$ or colorons predicted in TC2 model show up in upcoming collider experiments,
behavior of the model in low energy region for those discovered
particles can be tested and described by its effective electroweak
chiral Lagrangian (EWCL) \cite{EWCL,EWCLmatter} which, as a model
independent platform of investigating EWSB mechanism, parameterizes
the model by a set of coefficients. Starting from this EWCL, except
phenomenological research on fixing the coefficients of EWCL from
experiments data, theoretical studies concentrate on computing the
values of the coefficients from the detail underlying model.
Considering that TC2 model involves strongly coupled dynamics for
which traditional perturbative expansion fails in the computation
for the coefficients of EWCL, in previous paper\cite{TC2-07}, we
built up a formulation computing bosonic part of EWCL coefficients
up to order of $p^4$ for one-doublet TC model\cite{Farhi80} and
Hill's schematic model\cite{Hill95}. This formulation is of general
purposes, it can be applied to many other strongly coupled models.
Then EWCL becomes an universal platform on which we can compare
 different underlying models with experiment data and extract out the
 true physical theory of our real world. To achieve the aim of this
comparison, the left theoretical works are to compute EWCL
coefficients model by models. Present work is the second paper
starting from Ref.\cite{TC2-07} for series computations for various
strongly coupled new physics models. Here we focus on K.Lane's
prototype natural TC2 model\cite{Lane95}.

In original Hill's model, effects of ETC interactions are only
qualitatively estimated, effective four fermion interactions induced
by ETC (EFFIIETC) are even not explicitly written down in
Ref.\cite{Hill95}. Accordingly our previous computations
\cite{TC2-07} also do not involve possible ETC's contributions.
Examining ETC effects, Chivukula, Dobrescu and Terning
(CDT)\cite{CDT} argued that the TC2 proposal cannot be both natural
and consistent with experimental measurements of the parameter
$\rho=M_W^2/M_Z^2\cos^2\theta_W$. In extreme case, even for
degenerate up and down type technifermions of third generation are
likely to have custodial-isospin violating couplings to the strong
$U(1)_1$ since part of $m_t$ must arise from ETC, and this leads to
large contributions to $\rho$ parameter which contradicts with
experiment data. \footnote{In fact, the detail up and down type
technifermions of third generation are formally arranged not
participating $U(1)_1$ interaction by vanishing their $U(1)_1$
charges in original Hill's model and then do not cause large
contribution to $\rho$. This result is compatible with that obtained
in Ref.\cite{TC2-07}. But this naive arrangement is not realistic in
the sense, as mentioned by CDT\cite{CDT}, that to give top and
bottom (which must have different $U(1)_1$ charges to allow for
their different masses) ETC masses, the different right-handed
technifermions to which top and bottom quarks couple must have
different $U(1)_1$ charges. } To overcome this difficulty, instead
of conventional one doublet third generation technifermions, K.Lane
and E.Eichten propose their model\cite{Lane95} by introducing two
sets of technifermion doublets for third generation techniquarks
with different $U(1)_1$ charges but up and down type technifermions
in the same doublet possesing the same $U(1)_1$ charges:
$T^t_{L,R}=(U^t,D^t)_{L,R}$ giving the top quark its ETC-mass;
$T^b_{L,R}=(U^b,D^b)_{L,R}$ giving the bottom quark its ETC-mass,
these cut the intimate relation between custodial-isospin violation
from techniquarks and t-b mass difference. Due to this important
role of ETC interactions in Lane's model, its effects in EWCL is
worth of examination and this paper is not only for computing EWCL
coefficients of Lane's model, but also for investigating ETC effects
on these coefficients.

In next section, we apply our formulation developed in
Ref.\cite{TC2-07} to Lane's model\cite{Lane95}. We perform dynamical
calculations through several steps: first integrate in goldstone
field $U$, then integrate out technigluons and techniquarks by
solving Schwinger-Dyson equation (SDE) for techniquarks and compute
result effective action, further integrating out $Z'$ and finally
obtain EWCL coefficients. Section III is the discussion.
 In the appendix, we list some requisite formulae.

\section{Derivation of EWCL from Lane's Model}

Consider prototype natural TC2 model proposed by K.Lane and
E.Eichten\cite{Lane95}. The TC group is not specified in
Ref.\cite{Lane95}, but chosen to be $SU(N)$ in later Lane's improved
model\cite{Lane96}. For definiteness, we take $G_{\rm TC}=SU(N)$.
The gauge charge assignments of techniquarks in $G_{\rm TC}\otimes
SU(3)_1\otimes SU(3)_2\otimes SU(2)_L\otimes U(1)_{Y_1}\otimes
U(1)_{Y_2}$ are shown as Table I for which we choose the case B
solution\footnote{Case A solution, as mentioned by K.Lane in
Ref.\cite{Lane95}, would not be possible to generate proper ETC
masses for the t and b quarks and therefore not considered in this
work.\label{caseA}} to the anomaly conditions of Ref.\cite{Lane95}.

\begin{table}[h]
\small{{\bf TABLE I}.~Gauge charge assignments of techniquarks for
prototype natural TC2 model given in Ref.\cite{Lane95}. These
techniquarks are $SU(3)_1\otimes SU(3)_2$ singlets. }

\renewcommand{\arraystretch}{2}
\begin{tabular}{*{5}{c}}
\hline field & {\footnotesize$SU(N)$} &
{\footnotesize$SU(2)_L$} & {\footnotesize$U(1)_{\mbox{\tiny$Y_1$}}$} & {\footnotesize$U(1)_{\mbox{\tiny$Y_2$}}$}\\
\hline \hline $T_L^l$&N&2&0&0\\
$U_R^l$&N&1&0&$\frac{1}{2}$\\
$D_R^l$&N&1&0&-$\frac{1}{2}$\\
$T_L^t$&N&2&-1&1\\
$U_R^t$&N&1&-$\frac{1}{2}$&1\\
$D_R^t$&N&1&-$\frac{1}{2}$&0\\
$T_L^b$&N&2&1&-1\\
$U_R^b$&N&1&$\frac{1}{2}$&0\\
$D_R^b$&N&1&$\frac{1}{2}$&-1\\ \hline
\end{tabular}
\end{table}

The action of the symmetry breaking sector then is
\begin{eqnarray}
&&\hspace{-0.5cm}S_{\rm
SBS}[G_{\mu}^\alpha\!,A_{1\mu}^A\!,A_{2\mu}^A\!,
W_\mu^a\!,B_{1\mu}\!,B_{2\mu}\!,\bar{T}^l\!,T^l\!,\bar{T}^t\!,T^t\!,\bar{T}^b\!,T^b]\nonumber\\
&&\hspace{-0.5cm}=\int d^4x({\cal L}_{\rm gauge}+{\cal L}_{\rm
techniquark}+\mathcal{L}_\mathrm{breaking}+\mathcal{L}_\mathrm{4T})\;,~~\label{SBSdef}
\end{eqnarray}
with different part of Lagrangian given by
\begin{eqnarray}
&&\hspace{-0.5cm}{\cal
L}_\mathrm{gauge}=-\frac{1}{4}F_{\mu\nu}^\alpha F^{\alpha,\mu\nu}
-\frac{1}{4}A_{1\mu\nu}^A A_1^{A\mu\nu}-\frac{1}{4}A_{2\mu\nu}^A
A_2^{A\mu\nu}\nonumber\\
&&\hspace{0.8cm}-\frac{1}{4}W_{\mu\nu}^a
W^{a,\mu\nu}-\frac{1}{4}B_{1\mu\nu}
B^{1,\mu\nu}-\frac{1}{4}B_{2\mu\nu} B_2^{\mu\nu},~~
\end{eqnarray}
\begin{eqnarray}
&&\hspace{-0.5cm}{\cal L}_\mathrm{techniquark}\label{Ltechniquark}\\
&&\hspace{-0.5cm}=\bar{T}^{l}(i\slashed{\partial}-g_{\rm TC}t^\alpha
\slashed{G}^\alpha-g_{2}\frac{\tau^{a}}{2}\slashed{W}^{a}P_{L}-\frac{1}{2}q_{2}\slashed{B}_{2}\tau^{3}P_{R})T^{l}\nonumber\\
&&\hspace{0cm}+\bar{T}^{t}(i\slashed{\partial}-g_{\rm TC}t^\alpha
\slashed{G}^\alpha-g_{2}\frac{\tau^{a}}{2}\slashed{W}^{a}P_{L}+q_{1}\slashed{B}_{1}P_{L}\nonumber\\
&&\hspace{0cm}-q_{2}\slashed{B}_{2}P_{L}+\frac{1}{2}q_{1}\slashed{B}_{1}P_{R}-(\frac{1}{2}+\frac{\tau^{3}}{2})q_{2}\slashed{B}_{2}P_{R})T^{t}\nonumber\\
 &&\hspace{0cm}+\bar{T}^{b}(i\slashed{\partial}-g_{\rm
 TC}t^\alpha
\slashed{G}^\alpha-g_{2}\frac{\tau^{a}}{2}\slashed{W}^{a}P_{L}-q_{1}\slashed{B}_{1}P_{L}\nonumber\\
&&\hspace{0cm}+q_{2}\slashed{B}_{2}P_{L}-\frac{1}{2}q_{1}\slashed{B}_{1}P_{R}+(\frac{1}{2}-\frac{\tau^{3}}{2})q_{2}\slashed{B}_{2}P_{R})T^{b}\;,\nonumber\\
&&\hspace{-0.5cm}\mathcal
{L}_\mathrm{4T}=\mathcal{H}_\mathrm{diag}\;,\\
&&\hspace{-0.5cm}\mathcal{H}_\mathrm{diag}\!\!=\!\frac{g_\mathrm{ETC}^2}{M_\mathrm{ETC}^2}\bar{T}_L^i\gamma^{\mu}T_L^i
(b_U\bar{U}_R^j\gamma_{\mu}U_R^j+b_D\bar{D}_R^j\gamma_{\mu}D_R^j)\;,
\end{eqnarray}
where $g_\mathrm{TC}$, $g_2$, $q_1$ and $q_2$ are the coupling
constants of, respectively, $SU(N)$, $SU(2)_L$, $U(1)_{Y_1}$ and
$U(1)_{Y_2}$ (since techniquarks are $SU(3)_1\otimes SU(3)_2$
singlets, corresponding coupling constants do not show up here); and
the corresponding gauge fields (field strength tensors) are denoted
by $G_{\mu}^\alpha$, $W_\mu^a$, $B_{1\mu}$ and $B_{2\mu}$
($F_{\mu\nu}^\alpha$, $W_{\mu\nu}^a$, $B_{1\mu\nu}$ and
$B_{2\mu\nu}$) with the superscript $\alpha$ runs from 1 to $N^2-1$
 and $a$ from 1 to 3 ($SU(3)_1\otimes SU(3)_2$ gauge fields and field strength tensors
are denoted by $A_{1\mu}^A$, $A_{2\mu}^A$ and $A_{1\mu\nu}^A$,
$A_{2\mu\nu}^A$ with the superscript $A$ runs from 1 to 8);
$t^\alpha=\lambda^\alpha/2$ ($\alpha=1,\ldots,N^2-1$) and $\tau^a$
($a=1,2,3$) are, respectively, Gell-Mann and Pauli matrices.
$P_{^R_L}=(1\pm\gamma_5)/2$. Ordinary quarks are neglected, since we
only discuss bosonic part of EWCL.\footnote{For top quark, its
effect should be considered due to its large mass comparable to
symmetry breaking scale. There is an EFFIIETC
$\mathcal{H}_{\bar{t}t}=\frac{g_\mathrm{ETC}^2}{M_\mathrm{ETC}^2}\bar{t}_L\gamma^{\mu}U_L^t
\bar{U}_R^t\gamma_{\mu}t_R+h.c.$ responsible for top mass. This
interaction should be included in our calculation in principle and
if top quark has nonzero condensate, this interaction will
contribute to techniquark self energy. Since Ref.\cite{Lane95}
treats this term as a perturbation, we can ignore it at leading
order of our coefficients computations.} For ETC induced four
fermion interactions $\mathcal {L}_\mathrm{4T}$, although in
original Ref.\cite{Lane95}, except $\mathcal{H}_\mathrm{diag}$,
there other different kinds of interactions, such as
$\mathcal{H}_{\bar{l}t\bar{t}b}$ and $\mathcal
{H}_{\bar{l}b\bar{b}t}$, consider these non-diagonal interactions
will induce non-diagonal  condensates which violate the preferred
requirement
 $\langle\overline{U}^i_LU^j_R\rangle=\langle\overline{D}^i_LD^j_R\rangle\propto\delta_{ij}$
 for $i,j=l,t,b$ given in Ref.\cite{Lane95}, we drop them in our calculation.

 In Ref.\cite{Lane95}, an operator effecting $SU(3)_1\otimes
SU(3)_2\otimes U(1)_1\otimes U(1)_2$ breaking to $SU(3)_C\otimes
U(1)_Y$ is needed. We introduce a $3\times 3$ matrix scalar field
$\Phi$ to take the role of this operator to break $SU(3)_1\otimes
SU(3)_2\otimes U(1)_{Y_1}\otimes U(1)_{Y_2}$ to $SU(3)_C\otimes
U(1)_Y$ which leads massive colorons and $Z'$. This scalar field
transforms as $(\bar{3},3,\frac{5}{6},-\frac{5}{6})$ under the group
$SU(3)_1\otimes SU(3)_2\otimes U(1)_{Y_1}\otimes U(1)_{Y_{2}}$ which
leads covariant derivative
\begin{eqnarray}
D_{\mu}\Phi&=&\partial_{u}\Phi+i\Phi(h_1\frac{\lambda^{A*}}{2}A_{1\mu}^A
-\frac{5}{6}q_1B_{1\mu})\nonumber\\
&&-i(h_2\frac{\lambda^A}{2}A_{2\mu}^A
-\frac{5}{6}q_2B_{2\mu})\Phi\;,~~\nonumber
\end{eqnarray}
with $h_1$ and $h_2$ are the coupling constants of $SU(3)_1\otimes
SU(3)_2$ and corresponding Lagrangian can be written as
\begin{eqnarray}
\mathcal{L}_\mathrm{H}=\frac{1}{2}\mathrm{tr}[(D_{\mu}\Phi)^{\dagger}(D^{\mu}\Phi)]+V(\Phi)\label{LH}
\end{eqnarray}
in which potential $V(\Phi)$ is assumed to cause vacuum condensate
$\Phi_{ij}=v\delta_{ij}$ and the leading effects can be obtained by
just replacing $\Phi$ with its vacuum expectation value in
(\ref{LH}),
\begin{eqnarray}
\mathcal{L}_\mathrm{H}\!\!\stackrel{\Phi=v}{===}\!
\frac{1}{4}\frac{g_3^2}{\sin^2\!\theta\cos^2\!\theta}B_\mu^{A}B^{A\mu}\!
+\frac{25}{72}\frac{g_1^2}{\sin^2\!\theta'\cos^2\!\theta'}Z_\mu^{\prime}Z^{\prime\mu}\;,~~\label{LHv}
\end{eqnarray}
where the SM $U(1)_Y$ field $B_\mu$ with generator $Y=Y_1+Y_2$ and
the $U(1)^\prime$ field $Z_\mu^\prime$ (the gluon $A_\mu^A$ and
coloron $B_\mu^A$) are defined by orthogonal rotations with mixing
angle $\theta^\prime$ ($\theta$):
\begin{subequations}
\begin{eqnarray}
&&\begin{pmatrix}B_{1\mu} &
B_{2\mu}\end{pmatrix}=\begin{pmatrix}Z_\mu^\prime &
B_\mu\end{pmatrix}
\begin{pmatrix}\cos\theta^\prime & -\sin\theta^\prime\\ \sin\theta^\prime &
\cos\theta^\prime\end{pmatrix}\;,~~~~\label{B1B2-BZpri}\\
&&\begin{pmatrix}A_{1\mu}^A
& A_{2\mu}^A\end{pmatrix}=\begin{pmatrix}B_\mu^A &
A_\mu^A\end{pmatrix}
\begin{pmatrix}\cos\theta & -\sin\theta\\ \sin\theta &
\cos\theta\end{pmatrix}\;,~~~~\label{A1A2-AB}
\end{eqnarray}
\end{subequations}
with
\begin{eqnarray}
g_1\equiv
q_1\sin\theta^\prime=q_2\cos\theta^\prime\hspace{0.7cm}g_3\equiv
h_1\sin\theta=h_2\cos\theta\;.~~\label{g1-g2}
\end{eqnarray}
The coloron field $B^A_\mu$ does not couple to other fields except
to ordinary fermions at present order of approximation, so we can
ignore their contributions to bosonic part of EWCL. \footnote{One
can consider higher order corrections by including in (\ref{LHv})
the quantum fluctuation effects of field $\Phi$. Since these effects
depend on detail of symmetry breaking mechanism which is not
specified in Ref.\cite{Lane95}, in order not to deviate original
Lane's model too much, we ignore them in present paper.} i.e. we can
take
\begin{eqnarray}
\mathcal{L}_\mathrm{breaking}=\frac{1}{2}M_0^2Z_\mu^{\prime}Z^{\prime\mu}\hspace{0.5cm}
M_0^2=\frac{25}{36}\frac{g_1^2v^2}{\sin^2\!\theta'\cos^2\!\theta'}\;.~~\label{M0def}
\end{eqnarray}

With above preparations, the strategy to derive the EWCL from Lane's
model can be formulated as
\begin{eqnarray}
&&\hspace{-0.5cm}\exp\bigg(iS_{\mathrm{EW}}[W_\mu^a,B_\mu]\bigg)\\
&&\hspace{-0.5cm}=\int\mathcal{D}\bar{T}^l\mathcal{D}T^l\mathcal{D}\bar{T}^t\mathcal{D}T^t\mathcal{D}\bar{T}^b\mathcal{D}T^b
\mathcal{D}G_\mu^\alpha\mathcal{D}Z_\mu^\prime\exp\bigg[i\nonumber\\
&&\hspace{-0.2cm}\times S_{\mathrm{SBS}}[G_{\mu}^\alpha\!,0,0,
W_\mu^a\!,B_{1\mu}\!,B_{2\mu}\!,\bar{T}^l\!,T^l\!,\bar{T}^t\!,T^t\!,\bar{T}^b\!,T^b]\bigg]~~\label{strategy-TC20}\nonumber\\
&&\hspace{-0.5cm}=\mathcal{N}[W_\mu^a,B_\mu]\int\mathcal{D}\mu(U)\exp\bigg(iS_{\mathrm{eff}}[U,W_\mu^a,B_\mu]\bigg)\;,
\label{strategy-TC2}
\end{eqnarray}
where $A^A_\mu$ related to $A^A_{1\mu}$ and $A^A_{2\mu}$ through
(\ref{A1A2-AB}) is ordinary gluon field, $U(x)$ is a dimensionless
unitary unimodular matrix field in EWCL, and ${\cal D}\mu(U)$
denotes normalized functional integration measure on $U$. The
normalization factor $\mathcal{N}[W_\mu^a,B_\mu]$ is determined
through requirement that when the TC and ETC interactions are
switched off, $S_{\mathrm{eff}}[U,W_\mu^a,B_\mu]$ must vanishes.
This leads following electroweak gauge fields $W_\mu^a$, $B_\mu$
dependent $\mathcal{N}[W_\mu^a,B_\mu]$,
\begin{eqnarray}
&&\hspace{-0.5cm}\mathcal{N}[W_\mu^a,B_\mu]=\int\mathcal{D}\bar{T}^l\mathcal{D}T^l
\mathcal{D}\bar{T}^t\mathcal{D}T^t\mathcal{D}\bar{T}^b\mathcal{D}T^b
\mathcal{D}G_\mu^\alpha\mathcal{D}Z_\mu^\prime\nonumber\\
&&\hspace{1.7cm}\times~e^{iS_{\mathrm{SBS}}\big|_{\mathrm{ignore~TC,ETC},~A^A_{1\mu}=A^A_{2\mu}=0}}\;.~~~~
\label{Ndef}
\end{eqnarray}
Since there are many steps in deriving EWCL, in following several
subsections, we discuss them separately.
\subsection{Integrating in Goldstone Field $U$}

In terms of $Z'$ and $B$ fields given by (\ref{B1B2-BZpri}), we can
rewrite techniquark interaction (\ref{Ltechniquark}) as
\begin{eqnarray}
{\cal L}_\mathrm{techniquark} =\bar{\psi}(i\slashed{\partial}-g_{\rm
TC}t^\alpha
\slashed{G}^\alpha+\slashed{V}+\slashed{A}\gamma^{5})\psi\;,~~
\end{eqnarray}
where all three doublets techniquarks are arranged in one by six
matrix $\psi=(U^l,D^l,U^t,D^t,U^b,D^b)^T$ and
\begin{eqnarray}
&&V_\mu=(-\frac{1}{2}g_2\frac{\tau^a}{2}W_{\mu}^a
-\frac{1}{2}g_1\frac{\tau^3}{2}B_{\mu})\otimes\mathbf{I}+Z_{V\mu}\;,
\label{Vdef}\\
&&A_\mu=(\frac{1}{2}g_2\frac{\tau^a}{2}W_{\mu}^a
-\frac{1}{2}g_1\frac{\tau^3}{2}B_{\mu})\otimes\mathbf{I}+Z_{A\mu}\;,\label{Adef}
\end{eqnarray}
with $\mathbf{I}=\mathrm{diag}(1,1,1)$,
$Z_{V\mu}=\mathrm{diag}(Z_{V\mu}^l,Z_{V\mu}^t,Z_{V\mu}^b)$,
$Z_{A\mu}=\mathrm{diag}(Z_{A\mu}^l,Z_{A\mu}^t,Z_{A\mu}^b)$ and
\begin{eqnarray}
Z_{V\mu}^{l}&=&\frac{1}{4}g_1\tan\theta' Z^{\prime}_{\mu}\tau^3\;,\\
Z_{V\mu}^{t}&=&g_1Z^{\prime}_{\mu}[\frac{3}{4}\cot\theta'+(\frac{3}{4}+\frac{1}{4}\tau^3)\tan\theta']\;,\nonumber\\
Z_{V\mu}^{b}&=&g_1Z^{\prime}_{\mu}[-\frac{3}{4}\cot\theta'-(\frac{3}{4}-\frac{1}{4}\tau^3)\tan\theta']\;,\nonumber\\
Z_{A\mu}^{l}&=&\frac{1}{4}g_1\tan\theta' Z^{\prime}_{\mu}\tau^3\;,\label{ZAdef}\\
Z_{A\mu}^{t}&=&g_1Z^{\prime}_{\mu}[-\frac{1}{4}\cot\theta'+(-\frac{1}{4}+\frac{1}{4}\tau^3)\tan\theta']\;,\nonumber\\
Z_{A\mu}^{b}&=&g_1Z^{\prime}_{\mu}[\frac{1}{4}\cot\theta'+(\frac{1}{4}+\frac{1}{4}\tau^3)\tan\theta']\;.\nonumber
\end{eqnarray}

The Lagrangian (\ref{SBSdef}) is locally $SU(2)_{L}\times U(1)_{Y}$
invariant and approximately globally $SU(6)_{L} \times SU(6)_{R}$
invariant. We introduce a local $2\times2$ operator $O(x)$ as
$O(x)\equiv{\rm
tr}_{lc}[T^{l}_L(x)\bar{T}^{l}_R(x)+T^{t}_L(x)\bar{T}^{t}_R(x)+T^{b}_L(x)\bar{T}^{b}_R(x)]$
with ${\rm tr}_{lc}$ is the trace with respect to Lorentz and TC
indices. The transformation of $O(x)$ under $SU(2)_L\times U(1)_Y$
is $O(x)\rightarrow V_L(x)O(x)V_R^\dag(x)$ ( with
$V_L\!=\!e^{i\frac{\tau^a}{2}\theta^a}$ and
$V_R\!=\!e^{i\frac{\tau^3}{2}\theta^0}$). Then we decompose $O(x)$
as $O(x)=\xi_L^\dag(x)\sigma(x)\xi_R(x)$ with the $\sigma(x)$
represented by a hermitian matrix describes the modular degree of
freedom; while $\xi_L(x)$ and $\xi_R(x)$ are represented by unitary
matrices describe the phase degree of freedom of $SU(2)_L$ and
$U(1)_Y$ respectively.  Now we define a new field $U(x)$ as
$U(x)\equiv\xi_L^\dag(x)\xi_R(x)$ which
 is the nonlinear realization of the goldstone boson field in EWCL. Subtracting the $\sigma(x)$ field, we find that the present
decomposition results in a constraint
$\xi_L(x)O(x)\xi_R^\dag(x)-\xi_R(x)O^\dag(x)\xi_L^\dag(x)=0$, the
functional expression of it is
\begin{eqnarray}
\int\mathcal{D}\mu(U)\mathcal{F}[O]\delta(\xi_LO\xi_R^\dag-\xi_RO^\dag\xi_L^\dag)
=\mathrm{const.}\;,\label{InsrtU}
\end{eqnarray}
where $\mathcal{D}\mu(U)$ is an effective invariant integration
measure; $\mathcal{F}[O]$ only depends on $O$.  Substituting
identity (\ref{InsrtU}) into
 (\ref{strategy-TC20}), we obtain
\begin{eqnarray}
&&\hspace{-0.5cm}\int{\cal D}G_\mu^\alpha{\cal D}\bar{\psi}{\cal
D}\psi{\cal
D}Z^{\prime}_{\mu}\exp\bigg(\,iS_\mathrm{SBS}\big|_{A^A_{1\mu}=A^A_{2\mu}=0}\bigg)\nonumber\\
&&\hspace{-0.5cm}=\int\mathcal{D}\mu(U){\cal
D}Z^{\prime}_{\mu}\exp\bigg(iS_\mathrm{Z'}[U,W_\mu^a,B_\mu,Z^{\prime}_{\mu}]\bigg)\;,
\end{eqnarray}
where ${\cal D}\bar{\psi}{\cal D}\psi$ is the shorthand notation for
 $\mathcal{D}\bar{T}^l\mathcal{D}T^l
\mathcal{D}\bar{T}^t\mathcal{D}T^t\mathcal{D}\bar{T}^b\mathcal{D}T^b$
and
\begin{eqnarray}
&&\hspace{-0.5cm}S_\mathrm{Z'}[U,W_\mu^a,B_\mu,Z^{\prime}_{\mu}]\\
&=&\int
d^4x~(-\frac{1}{4}W_{\mu\nu}^aW^{a,\mu\nu}-\frac{1}{4}B_{\mu\nu}B^{\mu\nu}-\frac{1}{4}Z^{\prime}_{\mu\nu}Z^{\prime\mu\nu}\nonumber\\
&&+\frac{1}{2}M_0^2Z^{\prime}_{\mu}Z^{\prime\mu}) -i\log\int{\cal
D}G_\mu^\alpha{\cal D}\bar{\psi}{\cal
D}\psi\,\mathcal{F}[O]\nonumber\\
&&\times\delta(\xi_LO\xi_R^\dag-\xi_RO^\dag\xi_L^\dag)
\exp\bigg\{i\int d^4x[-\frac{1}{4}F_{\mu\nu}^\alpha
F^{\alpha,\mu\nu}\nonumber
\end{eqnarray}
\begin{eqnarray}
&&+\bar{\psi}(i\slashed{\partial}-g_{\rm TC}t^\alpha
\slashed{G}^\alpha+\slashed{V}+\slashed{A}\gamma^{5})\psi+\mathcal
{L}_{4T}]\bigg\}\;.\nonumber
\end{eqnarray}
From (\ref{strategy-TC2}), $S_\mathrm{eff}$ relates to
$S_\mathrm{Z'}$ by
\begin{eqnarray}
\mathcal{N}[W_\mu^a,B_\mu]e^{iS_\mathrm{eff}[U,W_\mu^a,B_\mu]}
=\!\int\mathcal{D}Z^{\prime}_{\mu}e^{iS_\mathrm{Z'}[U,W_\mu^a,B_\mu,Z^{\prime}_{\mu}]}~~~\label{SeffSZp}
\end{eqnarray}
 To match the correct normalization, we introduce in the
argument of logarithm function the normalization factor $\int{\cal
D}\bar{\psi}{\cal D}\psi e^{i\int
d^4x\bar{\psi}(i\slashed{\partial}+\slashed{V}+\slashed{A}\gamma^5)\psi}=
\exp\mathrm{Tr}\log(i\slashed{\partial}+\slashed{V}+\slashed{A}\gamma^5)$
and then take a special $SU(2)_L\times U(1)_Y$ rotation, as
$V_L(x)=\xi_L(x)$ and $V_R(x)=\xi_R(x)$, on both numerator and
denominator of the normalization factor
\begin{eqnarray}
&&\hspace{-0.5cm}S_\mathrm{Z'}[U,W_\mu^a,B_\mu,Z^{\prime}_{\mu}]\nonumber\\
&&\hspace{-0.5cm}=\int
d^4x~(-\frac{1}{4}W_{\mu\nu}^aW^{a,\mu\nu}-\frac{1}{4}B_{\mu\nu}B^{\mu\nu}
-\frac{1}{4}Z^{\prime}_{\mu\nu}Z^{\prime\mu\nu}\nonumber\\
&&+\frac{1}{2}M_0^2Z^{\prime}_{\mu}Z^{\prime\mu})-i\mathrm{Tr}\log(i\slashed{\partial}+\slashed{V}+\slashed{A}\gamma^{5})\nonumber\\
&&-i\log\frac{\int{\cal D}G_\mu^\alpha{\cal D}\bar{\psi}_\xi{\cal
D}\psi_\xi\,\mathcal{F}[O_\xi]\delta(O_\xi-O_\xi^\dag)e^{iS'}}
{\int{\cal D}\bar{\psi}_\xi{\cal D}\psi_\xi e^{iS'\big|_\mathrm{ignore~TC,ETC}}}\label{SZp}\\
&&\hspace{-0.5cm}S'= \int d^4x~[-\frac{1}{4}F_{\mu\nu}^\alpha
F^{\alpha,\mu\nu} +\bar{\psi}_\xi(i\slashed{\partial}-g_{\rm
TC}t^\alpha
\slashed{G}^\alpha\nonumber\\
&&\hspace{0.7cm}+\slashed{V}_\xi+\slashed{A}_\xi\gamma^5
)\psi_\xi+\mathcal {L}_{\xi4T}]\;,
\end{eqnarray}
where the rotated fields are denoted by subscript $\xi$ and they are
defined as follows
\begin{eqnarray}
&&\hspace{-0.5cm}T^{i}_\xi=P_L\xi_L(x)T^{i}_{L}(x)+P_R\xi_R(x)T^{i}_R(x)\,,i=l,t,b\nonumber\\
&&\hspace{-0.5cm}O_\xi(x)\equiv\xi_L(x)O(x)\xi_R^\dag(x)\hspace{0.5cm}
Z_{\xi,\mu}^{\prime}(x)\equiv
Z_{\mu}^{\prime}(x)\,,\label{ZpxiDef}\\
&&\hspace{-0.5cm}g_2\frac{\tau^a}{2}W_{\xi,\mu}^a(x)\equiv
\xi_L(x)[g_2\frac{\tau^a}{2}W_{\mu}^a(x)-i\partial_\mu]\xi_L^\dag(x)\\
&&\hspace{-0.5cm} g_1\frac{\tau^3}{2}B_{\xi,\mu}(x)\equiv
\xi_R(x)[g_1\frac{\tau^3}{2}B_{\mu}(x)-i\partial_\mu]\xi_R^\dag(x)\,.\label{BxiDef}
\end{eqnarray}
and $\mathcal {L}_{\xi4T}$ is $\mathcal {L}_{4T}$ with TC fields
replaced with rotated ones. It can be shown that
\begin{eqnarray}
\mathcal {L}_{\xi4T}=\mathcal {L}_{4T}\;.
\end{eqnarray}

Action (\ref{SZp}) can be further decomposed as
\begin{eqnarray}
&&\hspace{-0.5cm}S_\mathrm{Z'}[U,W_\mu^a,B_\mu,Z^{\prime}_{\mu}]\nonumber\\
&&\hspace{-0.5cm}=\int
d^4x~(-\frac{1}{4}W_{\mu\nu}^aW^{a,\mu\nu}-\frac{1}{4}B_{\mu\nu}B^{\mu\nu}
-\frac{1}{4}Z^{\prime}_{\mu\nu}Z^{\prime\mu\nu}\nonumber\\
&&\hspace{-0.2cm}+\frac{1}{2}M_0^2Z^{\prime}_{\mu}Z^{\prime\mu})+S_\mathrm{norm}[U,W_\mu^a,B_\mu,Z^{\prime}_{\mu}]\nonumber\\
&&\hspace{-0.2cm}+S_\mathrm{anom}[U,W_\mu^a,B_\mu,Z^{\prime}_{\mu}]\;,
\label{sefz}
\end{eqnarray}
where
\begin{eqnarray}
&&\hspace{-0.5cm}S_{\rm norm}[U,W_\mu^a,B_\mu]\nonumber\\
&&\hspace{-0.5cm}=-i\log\int{\cal D}G_\mu^\alpha{\cal
D}\bar{\psi}_\xi{\cal
D}\psi_\xi\,\mathcal{F}[O_\xi]\delta(O_\xi-O_\xi^\dag)~e^{iS'}\;,\label{action-eff2-norm}
\end{eqnarray}
and
\begin{eqnarray}
&&\hspace{-0.5cm}iS_{\rm
anom}[U,W_\mu^a,B_\mu,Z^{\prime}_{\mu}]\nonumber\\
&&\hspace{-0.5cm}=\mathrm{Tr}\log(i\slashed{\partial}+\slashed{V}+\slashed{A}\gamma^5)
-\mathrm{Tr}\log(i\slashed{\partial}+\slashed{V}_\xi+\slashed{A}_\xi\gamma^5)~\,.~~~
\end{eqnarray}
The transformations of the rotated fields under $SU(2)_L\times
U(1)_Y$ are $\psi_\xi(x)\rightarrow
h(x)\psi_\xi(x)$,~$O_\xi(x)\rightarrow h(x)O_\xi(x)h^\dag(x)$ with
$h(x)$ describes a hidden local $U(1)$ symmetry. Thus, the chiral
symmetry $SU(2)_L\otimes U(1)_Y$ covariance of the unrotated fields
has been transferred totally to the hidden symmetry $U(1)$
covariance of the rotated fields.
\subsection{Integrating out techinigluons and techniquarks}

With technique developed in Ref.\cite{TC2-07}, the integration over
technigulon fields in Eq.\eqref{action-eff2-norm} can be formally
integrated out with help of full $n$-point Green's function of the
$G_\mu^\alpha$-field
$G_{\mu_1\ldots\mu_n}^{\alpha_1\ldots\alpha_n}$,
\begin{eqnarray}
&&\hspace{-0.5cm}e^{iS_{\rm
norm}[U,W_\mu^a,B_\mu,Z^{\prime}_{\mu}]}\nonumber\\
&&\hspace{-0.5cm}=\int{\cal D}\bar{\psi}_\xi{\cal
D}\psi_\xi\,\mathcal{F}[O_\xi]\delta(O_\xi-O_\xi^\dag)\exp\bigg\{i\int
d^4x[\bar{\psi}_\xi(i\slashed{\partial}+\nonumber\\
&&\hspace{-0.2cm}\slashed{V}_\xi+\slashed{A}_\xi\gamma^5)\psi_\xi+\mathcal{L}_{\xi4T}]+\sum_{n=2}^\infty\int
d^4x_1\ldots d^4x_n\frac{(-ig_\mathrm{TC})^n}{n!}\nonumber\\
&&\hspace{-0.1cm}\times
G_{\mu_1\ldots\mu_n}^{\alpha_1\ldots\alpha_n}(x_1,\ldots,x_n)
J_{\xi,\alpha_1}^{\mu_1}(x_1)\ldots
J_{\xi,\alpha_n}^{\mu_n}(x_n)\bigg\}\;,\label{action-eff2-norm2}
\end{eqnarray}
where effective source $J_\xi^{\alpha\mu}(x)$ is identified as
$J_\xi^{\alpha\mu}(x)\equiv
\bar{\psi}_\xi(x)t^\alpha\gamma^\mu\psi_\xi(x)$.
\subsubsection{Schwinger-Dyson Equation for Techniquark Propagator}

To show that the TC interaction indeed induces the condensate
$\langle\bar{\psi}\psi\rangle\neq0$ which triggers EWSB, we
explicitly calculate the behavior of techniquark propagator
$S^{\sigma\rho}(x,x^\prime)\equiv
\langle\psi_\xi^\sigma(x)\bar{\psi}_\xi^\rho(x^\prime)\rangle$ in
the following. {\it Neglecting the factor
$\mathcal{F}[O_\xi]\delta(O_\xi-O_\xi^\dag)$}, the total functional
derivative of the integrand with respect to
$\bar{\psi}_\xi^{\sigma}(x)$ is zero,
\begin{eqnarray}
&&\hspace{-0.5cm}0=\int{\cal D}\bar{\psi}_\xi{\cal
D}\psi_\xi\frac{\delta}{\delta\bar{\psi}_\xi^\sigma(x)}\exp\bigg[\int
d^4x(\bar{\psi}_\xi I+\bar{I}\psi_\xi)+i\int d^4x[\nonumber\\
&&\bar{\psi}_\xi(i\slashed{\partial}+\slashed{V}_{\xi}+\slashed{A}_{\xi}\gamma^5)\psi_\xi+\mathcal{L}_{\xi4T}]
+\sum_{n=2}^\infty\int d^4x_1\ldots d^4x_n\nonumber\\
&&\frac{(-ig_{\rm
TC})^n}{n!}G_{\mu_1\ldots\mu_n}^{\alpha_1\ldots\alpha_n}(x_1,\ldots,x_n)
J_{\xi,\alpha_1}^{\mu_1}(x_1)\ldots
J_{\xi,\alpha_n}^{\mu_n}(x_n)\bigg]\;,\nonumber
\end{eqnarray}
where $I(x)$ and $\bar{I}(x)$ are the external sources for
techniquark fields, respectively, $\bar{\psi}_\xi(x)$ and
$\psi_\xi(x)$; and which leads to SDE for techniquark propagator,
\begin{eqnarray}
S_X(x,y)=\langle X(x)\bar{X}(y)\rangle\hspace{0.5cm}
X=U^l_{\xi},D^l_{\xi},U^t_{\xi},D^t_{\xi},U^b_{\xi},D^b_{\xi}\;.~~~
\end{eqnarray}
The detail derivation procedure is similar as that in
Ref.\cite{TC2-07}. The only difference is that now we have EFFIIETC
in the theory. The final obtained SDE is
\begin{eqnarray}
&&\hspace{-0.5cm}i\Sigma_X(x,y)=C_2(N)g_\mathrm{TC}^2
G_{\mu\nu}(x,y)[\gamma^{\mu}S_X(x,y)\gamma^{\nu}]\label{eq-SDE3}\\
&&-iC_X\gamma_{\mu}[P_LS_X(x,x)P_L
+P_RS_X(x,x)P_R]\gamma^{\mu}\delta(x-y)\;,\nonumber
\end{eqnarray}
with techniquark self energy defined as
\begin{eqnarray}
&&\hspace{-0.5cm}i\Sigma_X(x,y)\\
&&\hspace{-0.5cm}\equiv S^{-1}_X(x,y)
+i[i\slashed{\partial}_x+\slashed{V}_{\xi}(x)+\slashed{A}_{\xi}(x)\gamma_{5}]\delta(x-y)\;,\nonumber
\end{eqnarray}
and technigluon propagator
$G_{\mu\nu}^{\alpha\beta}(x,y)=\delta^{\alpha\beta}G_{\mu\nu}(x,y)$.
$C_2(N)=(N^2-1)/(2N)$ is Casimir operator from $(t^\alpha
t^\alpha)_{ab}=C_2(N)\delta_{ab}$ for the fundamental representation
of TC group $SU(N)$. Further $C_X$ is effective ETC induced four
fermion  coupling which is
\begin{eqnarray}
&&C_{U^l_\xi}=C_{U^t_\xi}=C_{U^b_\xi}=\frac{g^2_\mathrm{ETC}}{M^2_\mathrm{ETC}}b_U\label{CUdef}\\
&&C_{D^l_\xi}=C_{D^t_\xi}=C_{D^b_\xi}=\frac{g^2_\mathrm{ETC}}{M^2_\mathrm{ETC}}b_D\;.~~~\nonumber
\end{eqnarray}

In the following, we first consider the case of
$V_{\xi,\mu}=A_{\xi,\mu}=0$. In this situation, the technigluon
propagator in Landau gauge is
$G_{\mu\nu}^{\alpha\beta}(x,y)=\int\frac{d^4p}{(2\pi)^4}e^{-ip(x-y)}G_{\mu\nu}(p^2)$
 with $G_{\mu\nu}(p^2)=\frac{i}{-p^2[1+\Pi(-p^2)]}(g_{\mu\nu}-p_\mu
p_\nu/p^2)$. And the techniquark self energy and propagator are
respectively
\begin{eqnarray}
\left(\begin{array}{c}
\Sigma_X(x,y)\\
S_X(x,y)\end{array}\right)=\int\frac{d^4p}{(2\pi)^4}e^{-ip(x-y)}\times\left(\begin{array}{c}\Sigma_X(-p^2)\\
S_X(-p^2)\end{array}\right)\;,~~
\end{eqnarray}
with $S_X(p)=i/[\slashed{p}-\Sigma_X(-p^2)]$. Substitute above
results into the SDE and parameterize the technigluon propagator as
$\alpha_\mathrm{TC}[(p_E-q_E)^2]\equiv
g^2_\mathrm{TC}/(4\pi[1+\Pi(p^2_E)])$ for Euclidean momentum
$p_E,q_E$, we obtain following integration equation which with
angular approximation
$\alpha_\mathrm{TC}[(p_E-q_E)^2]=\alpha_\mathrm{TC}(p_E^2)\theta(p_E^2-q_E^2)
+\alpha_\mathrm{TC}(q_E^2)\theta(q_E^2-p_E^2)$, can be further
reduced to differential equation,
\begin{eqnarray}
&&\hspace{-0.5cm}i\Sigma_X(-p^2)=4\int\frac{d^4q}{(2\pi)^4}\bigg\{\frac{3\pi
C_2(N)\alpha_\mathrm{TC}[-(p-q)^2]}{(p-q)^2}+C_X\bigg\}\nonumber\\
&&\hspace{1.4cm}\times\bigg[\frac{\Sigma_X(-q^2)}{q^2-\Sigma_X^2(-q^2)}\bigg]\;.\label{SDeqJ=0}
\end{eqnarray}
Once above equation presents nonzero solution, we obtain nontrivial
techniquark condensate
\begin{eqnarray}
\langle\bar{X}(x)X'(x)\rangle=-4N\delta_{XX'}\int\frac{d^4p_E}{(2\pi)^4}\frac{\Sigma_X(p_E^2)}{p_E^2
+\Sigma_X^2(p_E^2)}\;,
\end{eqnarray}
which breaks $SU(2)_L\otimes U(1)_1\otimes U(1)_2$ to subgroup
$U(1)_\mathrm{em}$.

\begin{figure}[t]
\caption{Techniquark self energy $\Sigma(p_E^2)$.
$\Lambda_\mathrm{TC}$ is in unit of GeV and is fixed by $f=250$GeV.
} \hspace*{-4.5cm}\begin{minipage}[t]{\textwidth}\label{SigmaPic}
\includegraphics[scale=0.6]{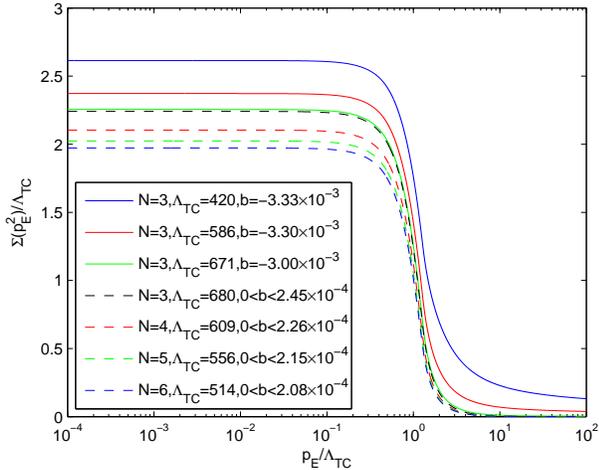}
\end{minipage}
\end{figure}

To obtain the numerical solution of equation (\ref{SDeqJ=0}), we
take the running constant $\alpha_\mathrm{TC}(p^2)$ the same as that
used in Eq.(49) of Ref.\cite{TC2-07} for which there are four input
parameters: $N,~N_f$, $\Lambda_\mathrm{TC}$ and $b$. $N$ as TC
number is a free parameter, we take four different values
$N=3,4,5,6$ estimating its effects; $N_f=6$ is due to three doublets
of techniquarks; The scale of TC interaction $\Lambda_\mathrm{TC}$
will be fixed from $f=250$GeV determined in later in
(\ref{f2beta1}); $b\equiv C\Lambda^2_\mathrm{TC}$ will be discussed
later with $C$ introduced in (\ref{CUdef}) as coefficients of
EFFIIETC.  We take the physical cutoff of the equation to be the
scale of ETC and
$\Lambda=\Lambda_\mathrm{ETC}=100\Lambda_\mathrm{TC}$. The result
$\Sigma(p^2_E)$ is depicted in Fig.\ref{SigmaPic} in which dashed
lines are for positive b and different $N$s; while solid lines are
for different negative $b$s and $N=3$.
From which, we find
\begin{enumerate}
\item For $N\!=\!3$ and positive $b$, EFFIIETC  infects $\Sigma(p_E^2)$ very little except to its large
momentum tail. We have changed coupling $b$ by enlarging its
magnitude 100 times, the general form of $\Sigma(p_E^2)$ almost do
not change. For $N=3$ and negative $b$, above $b=-0.00300$ the
change in $\Sigma(p_E^2)$ is small, below $b=-0.00300$, we see the
explicit change  of $\Sigma(p_E^2)$ which at large momentum region
exhibits typical slowly damping asymptotic behavior due to existence
of four fermion coupling. To check the validity of the phenomena, we
have changed differential equation to original integration equation
for SDE with and without angular approximation
$\alpha_\mathrm{TC}[(p_E-q_E)^2]=\alpha_\mathrm{TC}(p_E^2)\theta(p_E^2-q_E^2)
+\alpha_\mathrm{TC}(q_E^2)\theta(q_E^2-p_E^2)$ and increased the
cutoff of the theory, all obtain the similar result. For $N=4,5,6$,
we can find similar phenomena as the case of $N=3$ which are not
written down here, since later we will show that present model
prefers smaller $N$ and then the final result of our calculation
will be only limited in the case of $N=3$.
\item For large momentum tail of $\Sigma(p_E^2)$, we find that if positive $b$ is larger
than some critical value, $\Sigma(p_E^2)$ will be negative as
momentum becoming large which indicates the possible oscillation.
These values are $b_{N=3}=2.45\times 10^{-4}$,~$b_{N=4}=2.26\times
10^{-4}$,~$b_{N=5}=2.15\times 10^{-4}$,~$b_{N=6}=2.08\times
10^{-4}$. Considering that
$b\propto\Lambda^2_\mathrm{TC}/\Lambda^2_\mathrm{ETC}$ must be very
small, we take $b=2.08\times 10^{-4}$ as a typical value of our
computation. To exhibite the difference of tail for different $b$,
we draw diagrams of $\Sigma(p_E^2)$ with $b=2.08\times 10^{-4}$ and
$b=0$ together in Fig.\ref{SigmaPic1}.
\begin{figure}[t]
\caption{The tail of techniquark self energy $\Sigma(p_E^2)$
exhibits ETC effects. }
\hspace*{-4.5cm}\begin{minipage}[t]{\textwidth}\label{SigmaPic1}
\includegraphics[scale=0.6]{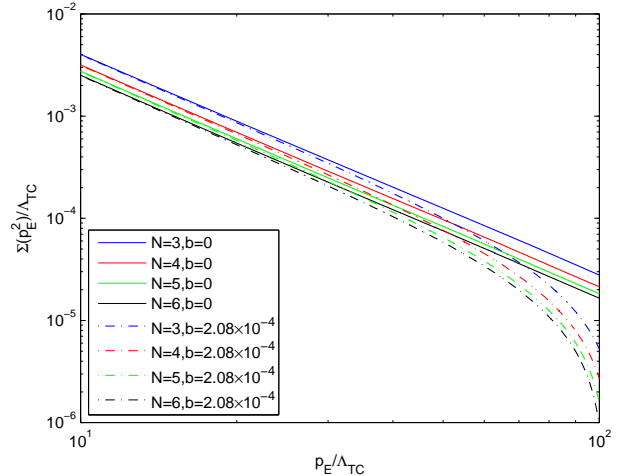}
\end{minipage}
\end{figure}
 We find that the differences show up only in the tail of self energy at momentum beyond
$50\Lambda_\mathrm{TC}$ and bellow that limit, there is almost no
difference. We further find that for fixed $f=250$GeV, from later
result of (\ref{f2beta1}), both $b=0$ and $b=2.08\times 10^{-4}$
cases all lead almost the same $\Lambda_\mathrm{TC}$.
\end{enumerate}

 If we further take $b_U=b_D$, $\Sigma_X$ equals for each techniflavor and  we
can neglect subscript $X$. Then with technique developed in
Ref.\cite{TC2-07}, we can show that if the function
$\Sigma(\partial_x^2)\delta(x-y)$ is the solution of the SDE in the
case $V_{\xi,\mu}=A_{\xi,\mu}=0$, we can replace its argument
$\partial_x$ by the minimal-coupling covariant derivative
$\overline{\nabla}_x\equiv\partial_x -iV_\xi(x)$ and use it, {\it
i.e.}, $\Sigma(\overline{\nabla}_x^2)\delta(x-y)$, as an approximate
solution of the SDE in the case $V_{\xi,\mu}\neq0$ and
$A_{\xi,\mu}\neq0$.
\subsubsection{Effective Action}

Starting from (\ref{action-eff2-norm2}),
 the exponential multi-fermion terms on the right-hand side of
 equation  can be written explicitly as
\begin{eqnarray}
&&\hspace{-0.5cm}\sum_{n=2}^\infty\int d^4x_1\ldots
d^4x_n\frac{(-ig_{\rm
TC})^n}{n!}G_{\mu_1\ldots\mu_n}^{\alpha_1\ldots\alpha_n}(x_1,\ldots,x_n)J_{\xi,\alpha_1}^{\mu_1}(x_1)\nonumber\\
&&\hspace{-0.5cm}\ldots
J_{\xi,\alpha_n}^{\mu_n}(x_n)\!\approx\!\!\int\! d^4xd^4x^\prime
\bar{\psi}_\xi^\sigma(x)\Pi_{\sigma\rho}(x,x^\prime)\psi_\xi^\rho(x^\prime),
\end{eqnarray}
\begin{eqnarray}
&&\hspace{-0.5cm}\Pi_{\sigma\rho}(x,x^\prime)=\sum_{n=2}^\infty\Pi_{\sigma\rho}^{(n)}(x,x^\prime)\approx
\Pi_{\sigma\rho}^2(x,x^\prime)\;,\\
&&\hspace{-0.5cm}\Pi_{\sigma\rho}^{(2)}(x,x^\prime)\!=\!-g_{\mathrm{TC}}^2G_{\mu_1\mu_2}^{\alpha_1\alpha_2}(x,x^\prime)
\bigg[t_{\alpha_1}\!\gamma^{\mu_1}\!S(x,x^{\prime})
t_{\alpha_2}\!\gamma^{\mu_2}\bigg]_{\sigma\rho}\;,~~~\label{eq-Pi-2}
\end{eqnarray}
where we have taken the approximation of {\it replacing the
summation over $2n$-fermion interactions with parts of them by their
vacuum expectation values (VEVs) and only keeping the leading four
fermion interactions.}  For $\mathcal {L}_{\xi4T}$ term in
(\ref{action-eff2-norm2}), we use the same approximation given
above.  Combining with result (\ref{eq-SDE3}) and neglecting the
factor $\mathcal{F}[O_\xi]\delta(O_\xi-O_\xi^\dag)$ in
Eq.\eqref{action-eff2-norm2}, we obtain
\begin{eqnarray}
&&\hspace{-0.5cm}S_{\rm
norm}[U,W_\mu^a,B_\mu]\nonumber\\
&&\hspace{-0.5cm}\approx-i\log\int{\cal D}\bar{\psi}_\xi{\cal
D}\psi_\xi\exp\bigg[i\int
d^4x\bar{\psi}_\xi(i\slashed{\partial}+\slashed{V}_{\xi}+\slashed{A}_{\xi}\gamma^5)\psi_\xi\nonumber\\
&&\hspace{-0.2cm}-i\int d^4xd^4x^\prime
~\bar{\psi}_\xi^\sigma(x)\Sigma_{\sigma\rho}(x,x^\prime)\psi_\xi^\rho(x^\prime)\bigg]\nonumber\\
&&\hspace{-0.5cm}\approx-i\mathrm{Tr}\log[i\slashed{\partial}+\slashed{V}_{\xi}+\slashed{A}_{\xi}\gamma^5
-\Sigma(\overline{\nabla}^2)]\;,
\end{eqnarray}
where $\Sigma(\overline{\nabla}^2)$ in techniflavor space is block
diagonal.  Notice that the arguments of $\mathrm{Tr}\log$ are block
diagonal which enable us to compute them block by blocks,
\begin{eqnarray}
&&\hspace{-0.5cm}S_\mathrm{norm}[U,W_\mu^a,B_\mu]\nonumber\\
&&\hspace{-0.5cm}=\sum_{\eta=1}^{3}-i\mathrm{Tr}\log[i\slashed{\partial}+\slashed{v}^{
\eta}+\slashed{a}^{\eta}\gamma^5-\Sigma(\overline{\nabla}^{\eta,2})]\nonumber\\
&&\hspace{-0.5cm}=\sum_{\eta=1}^{3}\int
d^4x\,\mathrm{tr}_f\bigg[(F_0^\mathrm{1D}
)^2a^{\eta2}-\mathcal{K}_1^\mathrm{1D}(d_\mu
a^{\eta\mu})^2\nonumber\\
&&\hspace{-0.2cm}-\mathcal{K}_2^\mathrm{1D}(d_\mu a_{\nu
}^{\eta}-d_\nu a_{\mu }^{\eta}
)^2+\mathcal{K}_3^\mathrm{1D}(a^{\eta2})^2+\mathcal{K}_4^\mathrm{1D}(a_{\mu }^{\eta} a_{\nu }^{\eta})^2\nonumber\\
&&-\mathcal{K}_{13}^\mathrm{1D}V_{\mu\nu }^{\eta}V^{\eta\mu\nu
}+i\mathcal{K}_{14}^\mathrm{1D}a_{\mu }^{\eta} a_{\nu  }^{\eta}
V^{\eta\mu\nu }\bigg]+\mathcal{O}(p^6)\;,\label{expand}
\end{eqnarray}
for which $\overline{\nabla}_\mu^\eta\equiv\partial_\mu
-iv_\xi^\eta$ and from (\ref{Vdef}) to (\ref{ZAdef}) and
(\ref{ZpxiDef}) to (\ref{BxiDef}),
\begin{eqnarray}
v^\eta_\mu&=&-\frac{1}{2}g_2\frac{\tau^a}{2}W_{\xi,\mu}^a
-\frac{1}{2}g_1\frac{\tau^3}{2}B_{\xi,\mu}+Z^\eta_{V\mu}\;,\label{vetaDef}\\
a^\eta_\mu&=&\frac{1}{2}g_2\frac{\tau^a}{2}W_{\xi,\mu}^a
-\frac{1}{2}g_1\frac{\tau^3}{2}B_{\xi,\mu}+Z^\eta_{A\mu}\hspace{1cm}\eta=l,t,b\;,\nonumber
\end{eqnarray}
where  $d_\eta a_{\nu }^\eta\equiv\partial_\mu
a_{\nu}^\eta-i[v_\mu^\eta, a_\nu^\eta]$,~$V_{\mu\nu}^\eta\equiv
i[\partial_\mu-iv_\mu^\eta,\partial_\nu-iv_\nu^\eta]$. $F_0^{1D}$
and $\mathcal{K}_i^\mathrm{1D}$ coefficients with superscript 1D to
denote that they are from one doublet TC model discussed in
Ref.\cite{TC2-07} which are functions of techniquark self energy
$\Sigma(p^2)$ and detailed expressions of them are already given in
 (36) of Ref.\cite{WQ02} with the replacement of $N_c\rightarrow
N$.

 For anomaly part, $U$ field dependent part
can be produced by normal part with vanishing techniquark self
energy $\Sigma$, i.e.
\begin{eqnarray}
&&\hspace{-0.5cm}iS_{\rm anom}[U,W_\mu^a,B_\mu]\\
&&\hspace{-0.5cm}=
\mathrm{Tr}\log(i\slashed{\partial}+\slashed{V}+\slashed{A}\gamma^5)
-iS_{\rm norm}[U,W_\mu^a,B_\mu]|_{\Sigma=0}\;.\nonumber
\end{eqnarray}
Notice that pure gauge field part independent of $U$ field is
irrelevant to EWCL. Combined with (\ref{expand}), above relation
imply
\begin{eqnarray}
&&\hspace{-0.5cm}iS_{\rm
anom}[U,W_\mu^a,B_\mu]\nonumber\\
&&\hspace{-0.5cm}=\mathrm{Tr}\log(i\slashed{\partial}\!+\!\slashed{V}\!+\!\slashed{A}\gamma^5)
\!+\!i\!\sum_{\eta=1}^{3}\!\!\int\!
d^4x\mathrm{tr}_f\!\bigg[\!-\mathcal{K}_1^\mathrm{1D,(anom)}(d_\mu
a^{\eta\mu})^2\nonumber\\
&&\hspace{-0.2cm}-\mathcal{K}_2^\mathrm{1D,(anom)}(d_\mu a_{\nu
}^\eta-d_\nu a_{\mu }^\eta)^2+\mathcal{K}_3^\mathrm{1D,(anom)}(a^{\eta2})^2\nonumber\\
&&\hspace{-0.2cm}+\mathcal{K}_4^\mathrm{1D,(anom)}(a_{\mu }^{\eta}
a_{\nu }^{\eta})^2-\mathcal{K}_{13}^\mathrm{1D,(anom)}V_{\mu
\nu}^{\eta}V^{\eta\mu\nu}\nonumber\\
&&\hspace{-0.2cm}+i\mathcal{K}_{14}^\mathrm{1D,(anom)}a_{\mu
}^{\eta} a_{\nu }^{\eta} V^{\eta\mu\nu}\bigg]+\mathcal{O}(p^6)\;,
\end{eqnarray}
with
\begin{eqnarray}
{\cal K}_i^\mathrm{1D,(anom)}=-{\cal
K}_i^\mathrm{1D}|_{\Sigma=0}\hspace{1cm}i=1,2,3,4,13,14~~~
\end{eqnarray}
where we have used result that $F_0^\mathrm{1D}|_{\Sigma=0}=0$.
Combining normal and anomaly part contributions together, with help
of (\ref{sefz}), we finally find
\begin{eqnarray}
&&\hspace{-0.5cm}S_\mathrm{Z'}[U,W_\mu^a,B_\mu,Z^{\prime}_{\mu}]\nonumber\\
&&\hspace{-0.5cm}=\int
d^4x(-\frac{1}{4}W_{\mu\nu}^aW^{a,\mu\nu}-\frac{1}{4}B_{\mu\nu}B^{\mu\nu}-\frac{1}{4}Z^{\prime}_{\mu\nu}Z^{\prime\mu\nu}\nonumber\\
&&\hspace{-0.2cm}+\frac{1}{2}M_0^2Z^{\prime}_{\mu}Z^{\prime\mu})
-i\mathrm{Tr}\log(i\slashed{\partial}+\slashed{V}+\slashed{A}\gamma^{5})+\sum_{\eta=1}^{3}\int d^4x\,\mathrm{tr}_f\bigg[\nonumber\\
&&\hspace{-0.2cm}(F_0^\mathrm{1D}
)^2a^{\eta2}-\mathcal{K}_1^{\mathrm{1D},\Sigma\neq0}(d_\mu
a^{\eta\mu})^2-\mathcal{K}_2^{\mathrm{1D},\Sigma\neq0}(d_\mu a_{\nu
}^{\eta}-d_\nu a_{\mu }^{\eta})^2\nonumber\\
&&\hspace{-0.2cm}+\mathcal{K}_3^{\mathrm{1D},\Sigma\neq0}(a^{\eta2})^2+\mathcal{K}_4^{\mathrm{1D},\Sigma\neq0}(a_{\mu
}^{\eta} a_{\nu}^{\eta})^2
-\mathcal{K}_{13}^{\mathrm{1D},\Sigma\neq0}V_{\mu\nu }^{\eta}V^{\eta\mu\nu}\nonumber\\
&&\hspace{-0.2cm}+i\mathcal{K}_{14}^{\mathrm{1D},\Sigma\neq0}a_{\mu}^{\eta}
a_{\nu }^{\eta} V^{\eta\mu\nu
}\bigg]+\mathcal{O}(p^6)\;.\label{allone}
\end{eqnarray}
With help of (\ref{vetaDef}) and (\ref{ZpxiDef}) to (\ref{BxiDef}),
above result can be further simplified to the form (\ref{Seff1}) in
which explicitly $U$ field dependence is displayed.
\subsection{Integrating out out $Z^{\prime}$}

We can further decompose (\ref{Seff1}) into
\begin{eqnarray}
&&\hspace{-0.5cm}S_\mathrm{Z'}[U,W_\mu^a,B_\mu,Z^{\prime}_{\mu}]\nonumber\\
&&\hspace{-0.5cm}=
\tilde{S}_\mathrm{Z'}[U,W_\mu^a,B_\mu,Z^{\prime}_{\mu}]+S_\mathrm{Z'}[U,W_\mu^a,B_\mu,0]\;,
\end{eqnarray}
where $\tilde{S}_\mathrm{Z'}[U,W_\mu^a,B_\mu,Z^{\prime}_{\mu}]$ is
$Z'$ dependent part of
$S_\mathrm{eff}[U,W_\mu^a,B_\mu,Z^{\prime}_{\mu}]$. We find $Z'$
independent part $S_\mathrm{Z'}[U,W_\mu^a,B_\mu,0]$ is just the same
as that of one-doublet TC model given in Ref.\cite{TC2-07}, the only
difference is that now there is an extra overall factor $3$
multiplied in front of all terms. The source of this factor comes
from the fact that in present model, instead of one doublet, we have
three techniquark doublets. So Switching off effects from $Z'$
particle, contributions of present TC2 model to bosonic part of EWCL
are equivalent to those of three-doublets TC model. In
$\tilde{S}_\mathrm{Z'}[U,W_\mu^a,B_\mu,Z^{\prime}_{\mu}]$, in order
to normalize $Z'$ field correctly, we introduce normalized field
$Z^{\prime}_{R,\mu}$ as
\begin{eqnarray}
&&\hspace{-0.5cm}Z^{\prime}_{\mu}=\frac{1}{c_{Z'}}Z^{\prime}_{R,\mu}\nonumber\\
&&\hspace{-0.5cm}c_{Z'}^2=1+g_1^2[3\mathcal{K}\tan^{2}\theta'
+10\mathcal{K}(\tan\theta'+\cot\theta')^{2}\label{cZ'def}\\
&&\hspace{0.3cm}+\mathcal{K}_2^{\mathrm{1D},\Sigma\neq0}(\tan\theta'+\cot\theta')^{2}
+\frac{3}{2}\mathcal{K}_{2}^{\mathrm{1D},\Sigma\neq0}\tan^{2}\theta'\nonumber\\
&&\hspace{0.3cm}+\frac{9}{2}\mathcal{K}_{13}^{\mathrm{1D},\Sigma\neq0}(\tan\theta'+\cot\theta')^{2}
+\frac{3}{2}\mathcal{K}_{13}^{\mathrm{1D},\Sigma\neq0}\tan^{2}\theta']\;,\nonumber
\end{eqnarray}
in terms of normalized field $Z^{\prime}_{R,\mu}$,
$\tilde{S}_\mathrm{Z'}[U,W_\mu^a,B_\mu,Z^{\prime}_{\mu}]$ become
\begin{eqnarray}
&&\hspace{-0.5cm}\tilde{S}_\mathrm{Z'}[U,W_\mu^a,B_\mu,Z^{\prime}_{\mu}]=\int
d^4x~[\frac{1}{2}Z'_{R,\mu}D_Z^{-1,\mu\nu}Z'_{R,\nu}~~~~\label{SZ'}\\
&&\hspace{2cm}+Z_R^{\prime,\mu}J_{Z,\mu}+Z_R^2Z_{R,\mu}'J^{\mu}_{3Z}
+g_{4Z}\frac{g_1^4}{c_{Z'}^{ 4}}Z_R^{\prime,4}]\nonumber
\end{eqnarray}
with
\begin{eqnarray}
&&\hspace{-0.5cm}D_Z^{-1,\mu\nu}=g^{\mu\nu}(\partial^2+M^2_{Z'})
-(1+\lambda_Z)\partial^{\mu}\partial^{\nu}+\Delta^{\mu\nu}_Z(X)\;,~~~~\label{DZdef}\\
&&\hspace{-0.5cm}M_{Z'}^2=\frac{1}{c_{Z'}^{2}}\{M_0^2+\frac{1}{2}(F_0^\mathrm{1D}
)^2g_1^2(\cot\theta'+\tan\theta')^2\nonumber\\
&&\hspace{0.6cm}+\frac{3}{4}(F_0^\mathrm{1D})^2g_1^2\tan^2\theta'\}\;, \label{MZdef}\\
&&\hspace{-0.5cm}\lambda_Z=\frac{g_1^2}{c_{Z'}^2}
[-\frac{1}{2}(\tan\theta'+\cot\theta')^{2}-\frac{3}{4}\tan^{2}\theta']\mathcal{K}_1^{\mathrm{1D},\Sigma\neq0}\;,\\
&&\hspace{-0.5cm}\Delta^{\mu\nu}_Z(X)=\frac{1}{c_{Z'}^{2}}\{(-\frac{3}{4}\mathcal{K}_{1}^{\mathrm{1D},\Sigma\neq0}
-\frac{3}{16}\mathcal{K}_{3}^{\mathrm{1D},\Sigma\neq0}
+\frac{3}{8}\mathcal{K}_{13}^{\mathrm{1D},\Sigma\neq0}\nonumber\\
&&\hspace{1cm}-\frac{3}{16}\mathcal{K}_{14}^{\mathrm{1D},\Sigma\neq0})g_{1}^{2}\tan^{2}\theta'
\mathrm{tr}[X^{\mu}\tau^{3}]\mathrm{tr}[X^{\nu}\tau^{3}]\nonumber\\
&&\hspace{1cm}+[\frac{3}{2}\mathcal{K}_1^{\mathrm{1D},\Sigma\neq0}\tan^2\theta'
-\frac{1}{4}(\cot\theta'+\tan\theta')^{2}\mathcal{K}_{3}^{\mathrm{1D},\Sigma\neq0}
\nonumber\\
&&\hspace{1cm}-\frac{1}{4}(\cot\theta'+\tan\theta')^{2}\mathcal{K}_{4}^{\mathrm{1D},\Sigma\neq0}
-\frac{3}{8}\mathcal{K}_4^{\mathrm{1D},\Sigma\neq0}\tan^2\theta'\nonumber\\
&&\hspace{0.5cm}-\frac{3}{4}\mathcal{K}_{13}^{\mathrm{1D},\Sigma\neq0}\tan^2\theta'
+\frac{3}{8}\mathcal{K}_{14}^{\mathrm{1D},\Sigma\neq0}\tan^2\theta']
g_1^2\mathrm{tr}[X^{\mu}X^{\nu}]\nonumber\\
&&\hspace{1cm}+g^{\mu\nu}[(-\frac{1}{8}(\cot\theta'+\tan\theta')^2
-\frac{3}{16}\tan^2\theta')\mathcal{K}_3^{\mathrm{1D},\Sigma\neq0}\nonumber\\
&&\hspace{1cm}+\frac{3}{16}\tan^2\theta'\mathcal{K}_4^{\mathrm{1D},\Sigma\neq0}
-\frac{1}{8}(\cot\theta'+\tan\theta')^{2}\mathcal{K}_4^{\mathrm{1D},\Sigma\neq0}\nonumber\\
&&\hspace{0.5cm}+\frac{3}{4}\tan^2\theta'\mathcal{K}_{13}^{\mathrm{1D},\Sigma\neq0}
-\frac{3}{8}\tan^2\theta'\mathcal{K}_{14}^{\mathrm{1D},\Sigma\neq0}]g_1^2\mathrm{tr}[X^{k}X_{k}]\nonumber\\
&&\hspace{0.5cm}+g^{\mu\nu}[-\frac{3}{16}\mathcal{K}_4^{\mathrm{1D},\Sigma\neq0}
-\frac{3}{8}\mathcal{K}_{13}^{\mathrm{1D},\Sigma\neq0}+\frac{3}{16}\mathcal{K}_{14}^{\mathrm{1D},\Sigma\neq0})]\nonumber\\
&&\hspace{0.5cm}\times
g_1^2\tan^2\theta'\mathrm{tr}[X_{k}\tau^{3}]\mathrm{tr}[X^{k}\tau^{3}]\}\;,\\
&&\hspace{-0.5cm}J_Z^\mu=J_{Z0}^\mu+\frac{g_1^2\gamma}{c_{Z'}}\partial^{\nu}B_{\mu\nu}+\tilde{J}_Z^\mu\;,\label{JZdef}\\
&&\hspace{-0.5cm}J_{Z0\mu}=-\frac{3}{4c_{Z'}}i(F_0^\mathrm{1D}
)^2g_1\tan\theta'\mathrm{tr}[X_{\mu}\tau^{3}]\;,\\
&&\hspace{-0.5cm}\gamma=3\mathcal{K}\tan\theta'+(\frac{3}{2}\mathcal{K}_2^{\mathrm{1D},\Sigma\neq0}
+\frac{3}{2}\mathcal{K}_{13}^{\mathrm{1D},\Sigma\neq0})\tan\theta'\;,\label{gammaDef}
\end{eqnarray}
\begin{eqnarray}
&&\hspace{-0.5cm}\tilde{J}_{Z}^\mu=\frac{1}{c_{Z'}}\bigg\{\frac{3}{4}ig_1\tan\theta'\mathcal{K}_1^{\mathrm{1D},\Sigma\neq0}\{
\mathrm{tr}[U^{\dag}(D^{\nu}D_{\nu}U)U^{\dag}D^{\mu}U\tau^{3}]\nonumber\\
&&\hspace{0.3cm}-\tan\theta'\mathrm{tr}[U^{\dag}(D^{\nu}D_{\nu}U)\tau^{3}U^{\dag}D^{\mu}U
+\partial^{\mu}(U^{\dag}D^{\nu}D_{\nu}U\tau^3)]\}\nonumber\\
&&\hspace{0.3cm}+\frac{3}{2}(-\mathcal{K}_2^{\mathrm{1D},\Sigma\neq0}
+\mathcal{K}_{13}^{\mathrm{1D},\Sigma\neq0})g_1\tan\theta'
\partial_{\nu}\mathrm{tr}[\overline{W}^{\mu\nu}\tau^3]\nonumber\\
&&\hspace{0.3cm}+\frac{3i}{4}(\frac{1}{4}\mathcal{K}_3^{\mathrm{1D},\Sigma\neq0}
-\frac{1}{4}\mathcal{K}_4^{\mathrm{1D},\Sigma\neq0}
-\mathcal{K}_{13}^{\mathrm{1D},\Sigma\neq0}+\frac{1}{2}\mathcal{K}_{14}^{\mathrm{1D},\Sigma\neq0})\nonumber\\
&&\hspace{0.3cm}\times
g_1\tan\theta'\mathrm{tr}[X^{\nu}X_{\nu}]tr[X^{\mu}\tau^3]+\frac{3i}{4}(
\frac{1}{2}\mathcal{K}_4^{\mathrm{1D},\Sigma\neq0}
\nonumber\\
&&\hspace{0.3cm}+\mathcal{K}_{13}^{\mathrm{1D},\Sigma\neq0}-\frac{1}{2}\mathcal{K}_{14}^{\mathrm{1D},\Sigma\neq0})g_1
\tan\theta'\mathrm{tr}[X^{\mu}X_{\nu}]\mathrm{tr}[X^{\nu}\tau^{3}]\nonumber\\
&&\hspace{-0.5cm}+\frac{3}{4}(-\mathcal{K}_{13}^{\mathrm{1D},\Sigma\neq0}
+\frac{1}{4}\mathcal{K}_{14}^{\mathrm{1D},\Sigma\neq0})g_1\tan\theta'
\mathrm{tr}[\overline{W}^{\mu\nu}(X_{\nu}\tau^3-\tau^3X_{\nu})]\nonumber\\
&&+\frac{3}{2}i(\mathcal{K}_{13}^{\mathrm{1D},\Sigma\neq0}-\frac{1}{4}\mathcal{K}_{14}^{\mathrm{1D},\Sigma\neq0})g_1\tan\theta'
\partial_\nu\mathrm{tr}[X^{\mu}X^\nu\tau^{3}]\bigg\}\;,\\
&&\hspace{-0.5cm}g_{4Z}=(\mathcal{K}_3^{\mathrm{1D},\Sigma\neq0}
+\mathcal{K}_4^{\mathrm{1D},\Sigma\neq0})[\frac{3}{128}\tan^{4}\theta'\label{g4Zdef}\\
&&\hspace{0.5cm}+\frac{3}{32}\tan^{2}\theta'(\cot\theta'+\tan\theta')^2+\frac{1}{64}(\cot\theta'+\tan\theta')^4]\;,\nonumber\\
&&\hspace{-0.5cm}J_{3Z}^\mu=\frac{-i}{c_{Z'}^{3}}(\mathcal{K}_3^{\mathrm{1D},\Sigma\neq0}
+\mathcal{K}_4^{\mathrm{1D},\Sigma\neq0})g_1^3
[\frac{3}{32}\tan^3\theta'\nonumber\\
&&\hspace{0.5cm}+\frac{3}{16}(\cot\theta'+\tan\theta')^2\tan\theta']\mathrm{tr}[X^{\mu}\tau_{3}]\;.
\end{eqnarray}
Perform loop expansion to (\ref{SeffSZp}), the result of $Z'$ field
integration is
\begin{eqnarray}
&&S_\mathrm{eff}[U,W_\mu^a,B_\mu]-i\log\mathcal{N}[W_\mu^a,B_\mu]\nonumber\\
&&=\tilde{S}_\mathrm{Z'}[Z'_c,U,W^a,B]+\mbox{loop terms}\;,
\end{eqnarray}
with classical field $Z'_c$ satisfy
\begin{eqnarray}
\frac{\partial }{\partial
Z'_{c,\mu}(x)}\bigg[\tilde{S}_\mathrm{Z'}[Z'_c,U,W^a,B]+\mbox{loop
terms}\bigg]=0\;,
\end{eqnarray}
and
\begin{eqnarray}
-i\log\mathcal{N}[W_\mu^a,B_\mu]=\bigg[\tilde{S}_\mathrm{Z'}[Z'_c,U,W^a,B]+\mbox{loop
terms}\bigg]_{\Sigma=0}\;,
\end{eqnarray}
which is obtained from (\ref{Ndef}) and the fact that when we switch
off TC and ETC interactions, techniquark self energy vanishes. With
(\ref{SZ'}), the solution is
\begin{eqnarray}
Z_c^{\prime\mu}(x)=-D^{\mu\nu}_ZJ_{Z,\nu}(x)+O(p^3)+\mbox{loop
terms}\;.
\end{eqnarray}
Then
\begin{eqnarray}
&&\hspace{-0.5cm}S_\mathrm{eff}[U,W_\mu^a,B_\mu]-i\log\mathcal{N}[W_\mu^a,B_\mu]\nonumber\\
&&\hspace{-0.5cm}=\int d^4x~[
-\frac{1}{2}J_{Z,\mu}D_Z^{\mu\nu}J_{Z,\nu}
-J_{3Z,\mu'}(D_Z^{\mu'\nu'}J_{Z,\nu'})(D_Z^{\mu\nu}J_{Z,\nu})^2\nonumber\\
&&+g_{4Z}\frac{g_1^4}{c_{Z'}}(D_Z^{\mu\nu}J_{Z,\nu})^4]+\mbox{loop
terms}\;,\label{SZ'out}
\end{eqnarray}
where
$D_Z^{-1,\mu\nu}D_{Z,\nu\lambda}=D_Z^{\mu\nu}D_{Z,\nu\lambda}^{-1}=g^\mu_\lambda$
and it is not difficult to show that if we are accurate up to order
of $p^4$, then  order $p$ of $Z_c'$ solution is enough, all
contributions from order $p^3$ of $Z'_c$ are at least belong to
order of $p^6$.
 With help of (\ref{SZ'out}), (\ref{DZdef}) and
(\ref{JZdef})
\begin{eqnarray}
&&\hspace{-0.5cm}S_\mathrm{eff}[U,W_\mu^a,B_\mu]-i\log\mathcal{N}[W_\mu^a,B_\mu]\nonumber\\
&&\hspace{-0.5cm}=\int
d^4x~[-\frac{1}{2}J_{Z0,\mu}D_Z^{\mu\nu}J_{Z0,\nu}
-\frac{1}{M_{Z'}^2}
J_{Z0,\mu}(\tilde{J}^\mu_Z+\frac{g_1^2\gamma}{c_{Z'}}\partial_{\nu}B^{\mu\nu})\nonumber\\
&&\hspace{-0.1cm}-\frac{1}{M_{Z'}^6}J_{3Z,\mu}J_{Z0}^{\mu}J_{Z0}^2
+\frac{g_{4Z}g_1^4}{c_{Z'}^{4}M_{Z'}^8}J_{Z0}^4]\;.\label{DeltaSZ'}
\end{eqnarray}
Ignoring terms higher than order of $p^4$, we find
$S_\mathrm{eff}[U,W_\mu^a,B_\mu]$ has exact form of standard EWCL up
to order of $p^4$. We can then read out the corresponding
coefficients, the result will be given in next subsection. The
normalization factor now is
\begin{eqnarray}
&&\hspace{-0.5cm}-i\log\mathcal{N}[W_\mu^a,B_\mu]=\int d^4x~[
-(\frac{1}{4}+\frac{3}{4}\mathcal{K}g_2^2+\frac{3}{8}\mathcal{K}_2^{\mathrm{1D},\Sigma\neq0}g_2^2\nonumber\\
&&\hspace{1cm}+\frac{3}{8}\mathcal{K}_{13}^{\mathrm{1D},\Sigma\neq0}g_2^2)W_{\mu\nu}^aW^{a,\mu\nu}
-(\frac{1}{4}+\frac{3}{4}\mathcal{K}g_1^2\nonumber\\
&&\hspace{1cm}+\frac{3}{8}\mathcal{K}_2^{\mathrm{1D},\Sigma\neq0}g_1^2
+\frac{3}{8}\mathcal{K}_{13}^{\mathrm{1D},\Sigma\neq0}g_1^2
+\frac{3(F_0^\mathrm{1D})^2}{8M_{Z^\prime}^2}\beta_1g_1^2\nonumber\\
&&\hspace{1cm}+\beta_1g_1^2\cot\theta'\gamma)B_{\mu\nu}B^{\mu\nu}]\;.
\end{eqnarray}
\subsection{Coefficients of EWCL}

\begin{figure}[t]
\caption{$S$ parameter for Lane's natural TC2 model.}\label{fig-S}
 \hspace*{-4.5cm}\begin{minipage}[t]{\textwidth}
    \includegraphics[scale=0.6]{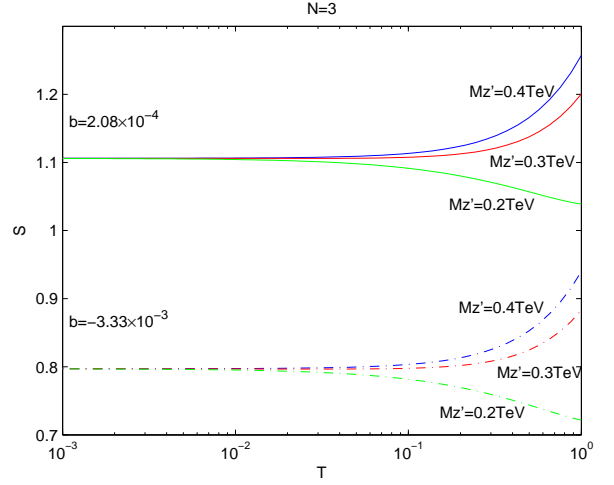}
\end{minipage}
\end{figure}

\begin{figure}[t]
\caption{$U$ parameter for Lane's natural TC2 model.}\label{fig-U}
 \hspace*{-4.5cm}\begin{minipage}[t]{\textwidth}
    \includegraphics[scale=0.6]{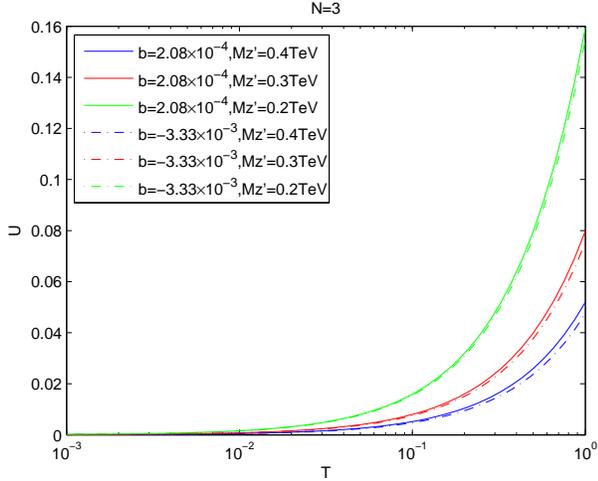}
\end{minipage}
\end{figure}

From $S_\mathrm{eff}[U,W_\mu^a,B_\mu]$ obtained in last subsection,
we can read out coefficients of EWCL. The $p^2$ order coefficients
are
\begin{eqnarray}
&&\hspace{-0.5cm}f^2=3(F_0^\mathrm{1D}
)^2\hspace{1cm}\beta_1=\frac{3(F_0^\mathrm{1D})^2g_1^2\tan^2\theta'}{8c_{Z'}^2M_{Z^\prime}^2}\;.
\label{f2beta1}
\end{eqnarray}
Combining with (\ref{M0def}),
(\ref{MZdef}) and $T$ parameter $\alpha T=2\beta_1$ given in
Ref.\cite{EWCL}, we further obtain
\begin{eqnarray}
\beta_1=\frac{1}{2}\alpha T=\frac{12}{ (\frac{200v^2}{3f^2}+
16)(1+\cot^2\theta')^2+24}\;,~~\label{betaTheta}
\end{eqnarray}
then $T$ is positive and uniquely determined by $\theta'$ and $v/f$.
It is bounded above and the upper limit is
$3/(5+25v^2/3f^2)\alpha\leq 9/(40\alpha)$, since we know $v\geq f$.
In following numerical computations, for simplicity, we all take
$v=f$. $p^4$ order coefficients are
\begin{eqnarray}
&&\hspace{-0.5cm}\alpha_1=3(1-2\beta_1)L^\mathrm{1D}_{10}+\frac{3(F_0^\mathrm{1D})^2}{2M_{Z^\prime}^2}\beta_1
-2\gamma\beta_1\cot\theta'\;,
\nonumber\\
&&\hspace{-0.5cm}\alpha_2=-\frac{3}{2}(1-2\beta_1)L^\mathrm{1D}_9
+\frac{3(F_0^\mathrm{1D})^2}{2M_{Z^\prime}^2}\beta_1
-2\gamma\beta_1\cot\theta'\;,\nonumber\\
&&\hspace{-0.5cm}\alpha_3=-\frac{3}{2}(1-2\beta_1)L^\mathrm{1D}_9\;,\nonumber\\
&&\hspace{-0.5cm}\alpha_4=3L^\mathrm{1D}_2+6\beta_1L^\mathrm{1D}_9
+\frac{3(F_0^\mathrm{1D})^2}{2M_{Z^\prime}^2}\beta_1\;,\nonumber\\
&&\hspace{-0.5cm}\alpha_5=3L^\mathrm{1D}_1+\frac{3}{2}L^\mathrm{1D}_3
-\frac{3(F_0^\mathrm{1D})^2}{2M_{Z^\prime}^2}\beta_1-6\beta_1L_9\;,\nonumber
\end{eqnarray}
\begin{eqnarray}
&&\hspace{-0.5cm}\alpha_6=-\frac{3(F_0^\mathrm{1D})^2}{2M_{Z^\prime}^2}\beta_1
-6\beta_{1}(4L^\mathrm{1D}_1+L^\mathrm{1D}_9)\nonumber\\
&&\hspace{0.3cm}+\beta_1^2[(1+\cot^2\theta')^2(48L^\mathrm{1D}_1+8L^\mathrm{1D}_3)+24L^\mathrm{1D}_1]\;,\nonumber\\
&&\hspace{-0.5cm}\alpha_7=\frac{3(F_0^\mathrm{1D})^2}{2M_{Z^\prime}^2}\beta_1
-2\beta_1(3L^\mathrm{1D}_3+6L^\mathrm{1D}_1-3L^\mathrm{1D}_9)+\beta_1^2[(1\nonumber\\
&&\hspace{0.3cm}+\cot^2\theta')^2(24L^\mathrm{1D}_1\!+4L^\mathrm{1D}_3)
+6\tan\theta'(L^\mathrm{1D}_3\!+2L^\mathrm{1D}_1)]\;,\nonumber\\
&&\hspace{-0.5cm}\alpha_8=-\frac{3(F_0^\mathrm{1D})^2}{2M_{Z^\prime}^2}\beta_1+12\beta_1L^\mathrm{1D}_{10}\;,\label{alpha8}\\
&&\hspace{-0.5cm}\alpha_9=-\frac{3(F_0^\mathrm{1D})^2}{2M_{Z^\prime}^2}\beta_1
+6\beta_1(L^\mathrm{1D}_{10}-L^\mathrm{1D}_9)\;,\nonumber\\
&&\hspace{-0.5cm}\alpha_{10}=4\beta_1^2(18L^\mathrm{1D}_1+3L^\mathrm{1D}_3)+32\beta_1^4g_{4Z}\cot^4\theta'\nonumber\\
&&\hspace{0.5cm}-\beta_1^3(144L^\mathrm{1D}_1+24L^\mathrm{1D}_3)[1+2(1+\cot^2\theta')^2]\;,\nonumber\\
&&\hspace{-0.5cm}\alpha_{11}=\alpha_{12}=\alpha_{13}=\alpha_{14}=0\;,\nonumber
\end{eqnarray}
where $L_i$ relate to $\mathcal{K}_i^{\mathrm{1D},\Sigma\neq0}$
coefficients through
\begin{eqnarray}
&&\hspace{-0.5cm}\mathcal{K}_2^{\mathrm{1D},\Sigma\neq0}=L^\mathrm{1D}_{10}-2H^\mathrm{1D}_1\;,\nonumber\\
&&\hspace{-0.5cm}\mathcal{K}_3^{\mathrm{1D},\Sigma\neq0}=64L^\mathrm{1D}_1+16L^\mathrm{1D}_3+8L^\mathrm{1D}_9
+2L^\mathrm{1D}_{10}+4H^\mathrm{1D}_1\;,\nonumber\\
&&\hspace{-0.5cm}\mathcal{K}_4^{\mathrm{1D},\Sigma\neq0}=32L^\mathrm{1D}_1-8L^\mathrm{1D}_9-2L^\mathrm{1D}_{10}
-4H^\mathrm{1D}_1\;,\nonumber\\
&&\hspace{-0.5cm}\mathcal{K}_{13}^{\mathrm{1D},\Sigma\neq0}=-L^\mathrm{1D}_{10}-2H^\mathrm{1D}_1\;,\nonumber\\
&&\hspace{-0.5cm}\mathcal{K}_{14}^{\mathrm{1D},\Sigma\neq0}=-4L^\mathrm{1D}_{10}-8L^\mathrm{1D}_9-8H^\mathrm{1D}_1\;.
\end{eqnarray}
 Several features of this result
are:
\begin{enumerate}
\item The contributions to $p^4$ order coefficients are divided into two parts:
a three doublets TC model contribution (equals to three times of one
doublet TC model discussed in Ref.\cite{TC2-07}) and $Z'$
contribution.
\item All corrections from $Z'$ particle are at
least proportional to $\beta_1$ which vanish if the mixing disappear
by $\theta'=0$.
\item Since $L_{10}^\mathrm{1D}<0$, combining with positive $\beta_1$, (\ref{alpha8}) then tells us $\alpha_8$ is
negative. Then $U=-16\pi\alpha_8$ coefficient given in
Ref.\cite{EWCL} is always positive in present model.
\item $\alpha_1$ and $\alpha_2$ depend on $\gamma$ which
from (\ref{gammaDef}) further rely on an extra parameter
$\mathcal{K}$. We can combine (\ref{f2beta1}) and (\ref{cZ'def})
together to fix $\mathcal{K}$,
\begin{eqnarray}
&&\hspace{0.2cm}\frac{(F_0^\mathrm{1D})^2g_1^2\tan^2\theta'}{8\beta_1M_{Z^\prime}^2}\label{cZ'def1}\\
&&\hspace{0.2cm}=\frac{1}{3}+g_1^2[\mathcal{K}\tan^{2}\theta'
+\frac{10}{3}\mathcal{K}(\tan\theta'+\cot\theta')^{2}\nonumber\\
&&\hspace{0.5cm}+\frac{1}{3}\mathcal{K}_2^{\mathrm{1D},\Sigma\neq0}(\tan\theta'+\cot\theta')^{2}
+\frac{1}{2}\mathcal{K}_{2}^{\mathrm{1D},\Sigma\neq0}\tan^{2}\theta'\nonumber\\
&&\hspace{0.5cm}+\frac{3}{2}\mathcal{K}_{13}^{\mathrm{1D},\Sigma\neq0}(\tan\theta'+\cot\theta')^{2}
+\frac{1}{2}\mathcal{K}_{13}^{\mathrm{1D},\Sigma\neq0}\tan^{2}\theta']\;.\nonumber
\end{eqnarray}
Once $\mathcal{K}$ is fixed, with help of (\ref{kappaDef}), we can
determine the ratio of infrared cutoff $\kappa$ and ultraviolet
cutoff $\Lambda$, in Fig.\ref{fig-cutoff1}, we draw the
$\kappa/\Lambda$ as function of $T$ and $M_{Z'}$, we  find natural
criteria $\Lambda>\kappa$ offers stringent constraints on the
allowed region for $T$ and $M_{Z'}$ that present theory prefer small
$Z'$ mass ($<0.4$TeV) and small TC group. For example, $T<0.035$ for
$M_{Z'}=0.3$TeV and $N=3$,  $T<0.25$ for $M_{Z'}=0.2$TeV and $N=6$,
$T<0.74$ for $M_{Z'}=0.2$TeV and $N=3$. In Fig.\ref{fig-cutoff2}, we
draw $Z'$ mass as function of $T$ parameter and $\kappa/\Lambda$.
\begin{figure}[t]
\caption{The ratio of infrared cutoff and ultraviolet cutoff
$\kappa/\Lambda$ as function of $T$ parameter and $Z'$ mass in unit
of TeV. } \label{fig-cutoff1}
\hspace*{-4.5cm}\begin{minipage}[t]{\textwidth}
    \includegraphics[scale=0.6]{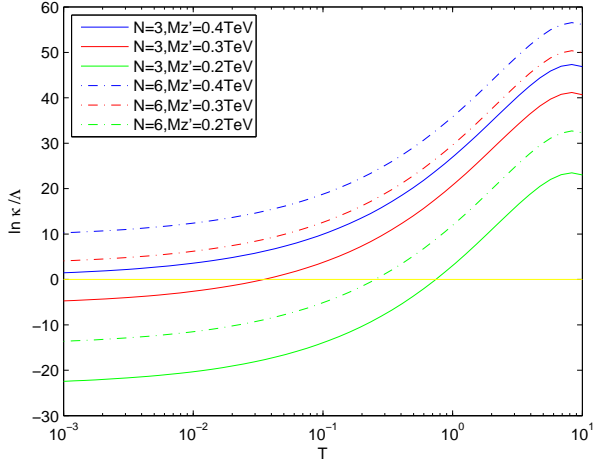}
\end{minipage}
\end{figure}
\begin{figure}[t]
\caption{ $Z'$ mass in unit of TeV as function of $T$ parameter and
$\kappa/\Lambda$.} \label{fig-cutoff2}
\hspace*{-4.5cm}\begin{minipage}[t]{\textwidth}
    \includegraphics[scale=0.6]{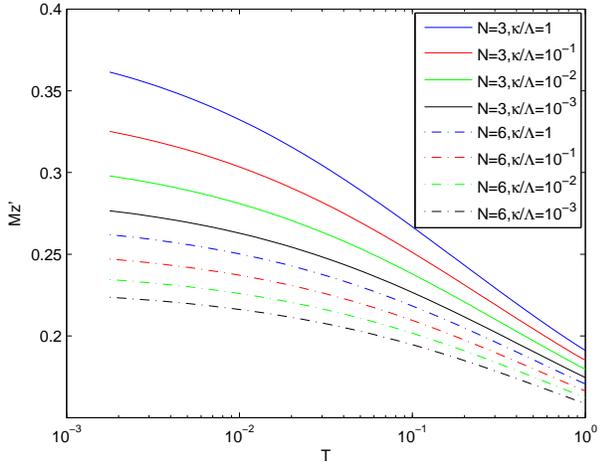}
\end{minipage}
\end{figure}
The line of $\kappa/\Lambda=1$ gives the upper bound of $Z'$ mass
$M_{Z'}<0.4$TeV, which is already beyond the experiment limit given
by Ref.\cite{CT,CS}. To check whether this bound is reliable, we
have changed  coupling of EFFIIETC by either enlarging its magnitude
100 times or reversing its sign, the results all almost do not
change. The special case of $b=-3.33\times 10^{-3}$ also has no
effects here. To examine the reason that why present model cause
smaller $M_{Z'}$ than that from Hill's model, we consider the
situation of very tiny $\theta'$ and $\kappa/\Lambda$, then the
leading term in r.h.s. of (\ref{cZ'def1}) is
$\frac{10}{3}g_1^2\mathcal{K}\cot^2\theta'$. Combining with
Eq.(\ref{betaTheta}), we find that (\ref{cZ'def1}) in this extreme
case gives $M_{Z'}=F_0\sqrt{\frac{31}{120}\mathcal{K}}\simeq
f\sqrt{\mathcal{K}}/(2\sqrt{3})$. while for Hill's model, we obtain
 result that $M_{Z'}=F_0\sqrt{\mathcal{K}}/2=f\sqrt{\mathcal{K}}/2$.
 So $Z'$ mass is smaller than that in Hill's
model by a factor $1/\sqrt{3}$ due to identification of $F_0$ with
$f/\sqrt{3}$ now in (\ref{f2beta1}) but with $f$ in Hill's model.
Considering smaller TC group will allow relative larger $Z'$ mass,
in following discussions, we only limit us in the case of $N=3$.
\item For typical case with $b=-3.33\times 10^{-3}$, except coefficients $F_0^\mathrm{1D}$ and
$\mathcal{K}_1^\mathrm{1D}$ which receive relative large corrections
from ETC interaction, all other $\mathcal{K}_i^\mathrm{1D}$
coefficients only feel small ETC effects.
\end{enumerate}
With $f=250$GeV, then  all EWCL coefficients depend on two physical
parameters $\beta_1$ and $M_{Z'}$. Combined with $\alpha
T=2\beta_1$, we can use the present experimental result for the $T$
parameter to fix $\beta_1$. In Fig.\ref{fig-S} and \ref{fig-U}, we
draw graphs for the $S=-16\pi\alpha_1$ and $U=-16\pi\alpha_8$ in
terms of the $T$ parameter respectively. We take three typical $Z'$
masses $M_{Z'}=0.2,~0.3,~0.4$TeV for references.
For $S$ parameter, we find that all values of it are at order of 1.
This can be understood as that at region of small $T$ parameter, the
main contribution to $S$ parameter comes from the three doublets TC
model which results in positive $S$, roughly equals to
$-3L_{10}^\mathrm{1D}$ which is three times larger than
corresponding value in Hill's model due to existence of three
doublets techniquarks. We also find that large negative $b$ will
reduce the value of $S$, but consider the value $b=-0.00333$
corresponding to
$g^2_\mathrm{ETC}b_U=-0.00333\Lambda^2_\mathrm{ETC}/\Lambda^2_\mathrm{TC}$
is already large enough, we do not expect more negative larger $b$
will have any physical meaning. For $U$ parameter, we find it is
positive and below $0.2$. Considering that the facts of small
$M_{Z'}$ and relative large $S$ are all not favored by present
precision measurements of SM, we just leave the analytic formulae
for other $\alpha_i$ coefficients there and will not draw diagrams
for them further more.

\section{Discussion}

 In this paper, we generalize the calculation in Ref.\cite{TC2-07} for
C.T.Hill's schematic TC2 model to K.Lane's prototype natural TC2
model. We find that, similar as Hill's model, coefficients of EWCL
for the Lane's model are divided into direct TC and ETC interaction
part, TC and topcolor induced effective $Z'$ particle contribution
part and ordinary quarks contribution part. The first two parts are
computed in this paper. We show that the direct TC and ETC
interaction part is three times larger than corresponding part of
Hill's model due to existence of three techniquark doublets , while
effective $Z'$ contributions are different with Hill's model due to
change of $U(1)_1\otimes U(1)_2$ group representation arrangements
and are at least proportional to the $p^2$ order parameter $\beta_1$
in EWCL. Typical features of the model are that it only allows
positive $S$, $T$ and $U$ parameters. $S$ is around $1$ which is
roughly three times larger than that in original Hill's model due to
existence of three doublets of techniquarks, and $T$ parameter
varies in the range $0\sim 9/(40\alpha)$. Analytical expression
(\ref{alpha8}) for five $p^4$ order coefficients including all three
custodial symmetry conserve ones
$\alpha_3,\alpha_4,\alpha_5,\alpha_8,\alpha_9$ exactly equal to
three times of those obtained from Hill's model in
Ref.\cite{TC2-07}. The $Z'$ mass is bounded from $0.4$TeV and larger
$M_{Z'}$ prefers smaller $N$. Compare to results obtained in
Ref.\cite{TC2-07} for C.T.Hill's TC2 model, the results from Lane's
first natural TC2 model deviate more from the experiment data. This
calls up for improvement of the model.

In fact, present model is only a prototype natural TC2 model. Many
details of the model are even not specified in original paper
\cite{Lane95} which prohibit us to perform computation more
accurately and leave us more space to improve the dynamics. One
typical non-specified effect is the walking dynamics. As metioned by
K.Lane that the TC of the model is expected to be a walking
 gauge theory.  This is of new feature different with conventional
 gauge theory, and this walking is not explicitly realized in present prototype model, since techniquarks
 are in fundamental representation of TC group and number of techniquarks is not large enough.
 Another unspecified detail is the $SU(3)_1\otimes SU(3)_2$ symmetry breaking
 mechanism. It now is simulated without detail
dynamics content by introducing an effective scalar field $\Phi$
which transforms as $(\bar{3},3,\frac{5}{6},-\frac{5}{6})$ under the
group $SU(3)_1\otimes SU(3)_2\otimes U(1)_{Y_1}\otimes U(1)_{Y_{2}}$
and corresponding interaction potential $V(\Phi)$. Introducing
scalar fields, although now only is effective, deviates the basic
idea of TC models. All these shortcomings are overcome in late
improved model \cite{Lane96}. Considering that this new model is
much more complex and different than present one which involves more
of different dynamics and then requires more analysis and
computation techniques. For example, the condensates of the
techniquarks are block diagonal in three doublets flavor space now
but not in the improved model (which is more like the case A
solution of the paper \cite{Lane95}, while present paper we only
discuss the case B solution and ignore case A as mentioned in
footnote $^{\ref{caseA}}$). In order to make our discussion not too
much complex and specially exhibit the result for Lane's first
natural TC2 model, in this paper, we only limit ourself in the
primary prototype model and focus our attention on figuring out
analytical expression for coefficients of EWCL, estimating possible
constrains to the model and identifying the effects of ETC
interactions, we leave the discussion of that new improved model in
future paper.
\section*{Acknowledgments}
This work was  supported by National  Science Foundation of China
(NSFC) under Grant No. 10435040 and No.10875065.

\appendix
\section{Necessary formulae for EWCL}

In this appendix, we list down the necessary formulae needed in the
text. With definition in which
\begin{eqnarray}
D_{\mu}U&=&\partial_{\mu}U+ig_{2}\frac{\tau^a}{2}W_{\mu}^aU-ig_{1}U\frac{\tau^3}{2}B_{\mu}\;,\\
D_{\mu}U^{\dag}&=&\partial_{\mu}U^{\dag}-ig_{2}U^{\dag}\frac{\tau^a}{2}W_{\mu}^a+ig_{1}\frac{\tau^3}{2}B_{\mu}U^{\dag}\;,\\
X_{\mu}&=&U^{\dag}(D_{\mu}U)\hspace{0.5cm}
\overline{W}_{\mu\nu}=U^{\dag}g_{2}\frac{\tau^{a}}{2}W^{a}_{\mu\nu}U\;.
\end{eqnarray}
we have
\begin{eqnarray}
&&\hspace{-0.5cm}S_\mathrm{Z'}[U,W_\mu^a,B_\mu,Z^{\prime}_{\mu}]\label{Seff1}\nonumber\\
&&\hspace{-0.5cm}=\int
d^4x\bigg\{-\frac{1}{4}W_{\mu\nu}^aW^{a,\mu\nu}-\frac{1}{4}B_{\mu\nu}B^{\mu\nu}-\frac{1}{4}Z^{\prime}_{\mu\nu}Z^{\prime\mu\nu}
+\frac{1}{2}M_0^2Z^{\prime}_{\mu}Z^{\prime\mu}
-\mathcal{K}~[\frac{3}{4}g_1^2B_{\mu\nu}B^{\mu\nu}-\frac{3}{2}g_1^2\tan\theta'
B_{\mu\nu}Z^{\prime\mu\nu}\nonumber\\
&&\hspace{-0.2cm}+\frac{3}{4}g_1^2\tan^2\theta'
Z^{\prime}_{\mu\nu}Z^{\prime\mu\nu}+\frac{5}{2}g_1^2(\tan\theta'+\cot\theta')^2Z^{\prime}_{\mu\nu}Z^{\prime\mu\nu}
+\frac{3}{4}g_2^2W^a_{\mu\nu}W^{a\mu\nu}]+(F_0^\mathrm{1D})^2\{-\frac{3}{4}\mathrm{tr}[X^{\mu}X_{\mu}]\nonumber\\
&&\hspace{-0.2cm}+\frac{1}{4}g_1^2(\cot\theta'+\tan\theta')^{2}Z^{\prime
2}+\frac{3}{8}g_1^2\tan^2\theta' Z^{\prime
2}-i\frac{3}{4}g_1\tan\theta'
Z^{\prime\mu}\mathrm{tr}[X_{\mu}\tau^{3}]\}
-\mathcal{K}_1^{\mathrm{1D},\Sigma\neq0}\{-\frac{3}{4}\mathrm{tr}[U^{\dag}(D^{\mu}D_{\mu}U)U^{\dag}(D^{\nu}D_{\nu}U)
\nonumber\\
&&\hspace{-0.2cm}+2U^{\dag}(D^{\mu}D_{\mu}U)(D^{\nu}U^{\dag})(D_{\nu}U)]
-\frac{3}{4}ig_1\tan\theta'Z^{\prime\nu}\mathrm{tr}[U^{\dag}(D^{\mu}D_{\mu}U)U^{\dag}D_{\nu}U\tau^{3}]
+\frac{3}{4}ig_1\tan\theta' Z^{\prime\nu}\mathrm{tr}[U^{\dag}(D^{\mu}D_{\mu}U)\tau^{3}U^{\dag}D_{\nu}U]\nonumber\\
&&\hspace{-0.2cm}-\frac{3}{4}ig_1\tan\theta'\partial_{\nu}Z^{\prime\nu}\mathrm{tr}[U^{\dag}(D^{\mu}D_{\mu}U)\tau^3]\}
+\frac{3}{8}(\mathcal{K}_1^{\mathrm{1D},\Sigma\neq0}+\frac{1}{4}\mathcal{K}_3^{\mathrm{1D},\Sigma\neq0}
-\frac{1}{4}\mathcal{K}_4^{\mathrm{1D},\Sigma\neq0}-\mathcal{K}_{13}^{\mathrm{1D},\Sigma\neq0}
+\frac{1}{2}\mathcal{K}_{14}^{\mathrm{1D},\Sigma\neq0})\nonumber\\
&&\hspace{-0.2cm}\times[\mathrm{tr}(X^{\mu}X_{\mu})]^{2}
-\frac{3}{8}(\mathcal{K}_1^{\mathrm{1D},\Sigma\neq0}+\frac{1}{4}\mathcal{K}_3^{\mathrm{1D},\Sigma\neq0}
-\frac{1}{2}\mathcal{K}_{13}^{\mathrm{1D},\Sigma\neq0}
+\frac{1}{4}\mathcal{K}_{14}^{\mathrm{1D},\Sigma\neq0})g_1^2\tan^2\theta'
Z^{\prime\mu}Z^{\prime}_\nu\mathrm{tr}[X_{\mu}\tau^{3}]\mathrm{tr}[X^{\nu}\tau^{3}]\nonumber\\
&&\hspace{-0.2cm}+\frac{1}{8}[6\mathcal{K}_1^{\mathrm{1D},\Sigma\neq0}\tan^2\theta'
-(\cot\theta'+\tan\theta')^2\mathcal{K}_3^{\mathrm{1D},\Sigma\neq0}
-(\cot\theta'+\tan\theta')^{2}\mathcal{K}_4^{\mathrm{1D},\Sigma\neq0}
-\frac{3}{2}\mathcal{K}_4^{\mathrm{1D},\Sigma\neq0}\tan^2\theta'\nonumber\\
&&\hspace{-0.2cm}-3\mathcal{K}_{13}^{\mathrm{1D},\Sigma\neq0}\tan^2\theta'
+\frac{3}{2}\mathcal{K}_{14}^{\mathrm{1D},\Sigma\neq0}\tan^2\theta']
g_1^2Z^{\prime\mu}Z^{\prime}_\nu\mathrm{tr}[X_{\mu}X^{\nu}]
+[-\frac{1}{4}(\tan\theta'+\cot\theta')^2-\frac{3}{8}\tan^2\theta']\mathcal{K}_1^{\mathrm{1D},\Sigma\neq0}
g_1^2(\partial_{\mu}Z^{\prime\mu})^2\nonumber\\
&&\hspace{-0.2cm}-\frac{3}{8}(\mathcal{K}_2^{\mathrm{1D},\Sigma\neq0}
+\mathcal{K}_{13}^{\mathrm{1D},\Sigma\neq0})(g_2^2W^{\mu\nu
a}W_{\mu\nu
a}+g_1^2B^{\mu\nu}B_{\mu\nu})+\frac{3}{4}(\mathcal{K}_2^{\mathrm{1D},\Sigma\neq0}
-\mathcal{K}_{13}^{\mathrm{1D},\Sigma\neq0})g_1\mathrm{tr}[\overline{W}^{\mu\nu}\tau^3](B_{\mu\nu}
-\tan\theta'Z^{\prime}_{\mu\nu})\nonumber\\
&&\hspace{-0.2cm}+\frac{3}{4}(\mathcal{K}_2^{\mathrm{1D},\Sigma\neq0}
+\mathcal{K}_{13}^{\mathrm{1D},\Sigma\neq0})\tan\theta'g_1^2Z^{\prime}_{\mu\nu}B^{\mu\nu}
-\frac{1}{4}[\mathcal{K}_2^{\mathrm{1D},\Sigma\neq0}(\tan\theta'+\cot\theta')^2
+\frac{3}{2}\mathcal{K}_2^{\mathrm{1D},\Sigma\neq0}\tan^2\theta'\nonumber\\
&&\hspace{-0.2cm}+9\mathcal{K}_{13}^{\mathrm{1D},\Sigma\neq0}(\tan\theta'+\cot\theta')^2
+\frac{3}{2}\mathcal{K}_{13}^{\mathrm{1D},\Sigma\neq0}\tan^2\theta']g_1^2Z^{\prime}_{\mu\nu}Z^{\prime\mu\nu}
+\frac{3i}{4}(\frac{1}{4}\mathcal{K}_3^{\mathrm{1D},\Sigma\neq0}-\frac{1}{4}\mathcal{K}_4^{\mathrm{1D},\Sigma\neq0}
-\mathcal{K}_{13}^{\mathrm{1D},\Sigma\neq0}\nonumber\\
&&\hspace{-0.2cm}+\frac{1}{2}\mathcal{K}_{14}^{\mathrm{1D},\Sigma\neq0})g_1\tan\theta'
Z^{\prime}_{\nu}\mathrm{tr}[X^{\mu}X_{\mu}]\mathrm{tr}[X^{\nu}\tau^{3}]
+\frac{1}{8}[-\frac{1}{2}(\cot\theta'+\tan\theta')^2\mathcal{K}_3^{\mathrm{1D},\Sigma\neq0}
-\frac{3}{4}\tan^2\theta'\mathcal{K}_3^{\mathrm{1D},\Sigma\neq0}\nonumber\\
&&\hspace{-0.2cm}+\frac{3}{4}\tan^2\theta'\mathcal{K}_4^{\mathrm{1D},\Sigma\neq0}
-\frac{1}{2}(\cot\theta'+\tan\theta')^2\mathcal{K}_4^{\mathrm{1D},\Sigma\neq0}
+3\tan^2\theta'\mathcal{K}_{13}^{\mathrm{1D},\Sigma\neq0}
-\frac{3}{2}\tan^2\theta'\mathcal{K}_{14}^{\mathrm{1D},\Sigma\neq0}]g_1^2Z^{\prime
2}\mathrm{tr}[X^{\mu}X_{\mu}]\nonumber\\
&&\hspace{-0.2cm}-\frac{3}{16}(\mathcal{K}_3^{\mathrm{1D},\Sigma\neq0}
+\mathcal{K}_4^{\mathrm{1D},\Sigma\neq0})ig_1^3[\frac{1}{2}\tan^3\theta'
+(\cot\theta'+\tan\theta')^2\tan\theta']Z^{\prime}_{\mu}Z^{\prime
2}\mathrm{tr}[X^{\mu}\tau_3]+\frac{1}{64}(\mathcal{K}_3^{\mathrm{1D},\Sigma\neq0}
+\mathcal{K}_4^{\mathrm{1D},\Sigma\neq0})g_1^4\nonumber\\
&&\hspace{-0.2cm}\times[\frac{3}{2}\tan^4\theta'+6\tan^2\theta(\cot\theta'+\tan\theta')^2
+(\cot\theta'+\tan\theta')^{4}]Z^{\prime
4}+\frac{3}{8}(\frac{1}{2}\mathcal{K}_4^{\mathrm{1D},\Sigma\neq0}+\mathcal{K}_{13}^{\mathrm{1D},\Sigma\neq0}
-\frac{1}{2}\mathcal{K}_{14}^{\mathrm{1D},\Sigma\neq0})\nonumber\\
&&\hspace{-0.2cm}\times\mathrm{tr}[X^{\mu}X_{\nu}]\mathrm{tr}[X_{\mu}X^{\nu}]
+\frac{3i}{4}(\frac{1}{2}\mathcal{K}_4^{\mathrm{1D},\Sigma\neq0}+\mathcal{K}_{13}^{\mathrm{1D},\Sigma\neq0}
-\frac{1}{2}\mathcal{K}_{14}^{\mathrm{1D},\Sigma\neq0})g_1\tan\theta'
Z^{\prime\nu}\mathrm{tr}[X_{\mu}X_{\nu}]
\mathrm{tr}[X^{\mu}\tau^3]\nonumber\\
&&\hspace{-0.2cm}+\frac{3}{16}(-\frac{1}{2}\mathcal{K}_4^{\mathrm{1D},\Sigma\neq0}
-\mathcal{K}_{13}^{\mathrm{1D},\Sigma\neq0}
+\frac{1}{2}\mathcal{K}_{14}^{\mathrm{1D},\Sigma\neq0})g_1^2\tan^2\theta'
Z^{\prime2}\mathrm{tr}[X_{\mu}\tau^{3}]\mathrm{tr}[X^{\mu}\tau^3]
+\frac{3i}{4}(-\mathcal{K}_{13}^{\mathrm{1D},\Sigma\neq0}
+\frac{1}{4}\mathcal{K}_{14}^{1D,\Sigma\neq0})g_1B_{\mu\nu}\nonumber\\
&&\hspace{-0.2cm}\times\mathrm{tr}[\tau^3X^{\mu}X^{\nu}]
+\frac{3i}{2}(-\mathcal{K}_{13}^{\mathrm{1D},\Sigma\neq0}+\frac{1}{4}\mathcal{K}_{14}^{\mathrm{1D},\Sigma\neq0})
\mathrm{tr}[X^{\mu}X^{\nu}\overline{W}_{\mu\nu}]
+\frac{3}{4}(-\mathcal{K}_{13}^{\mathrm{1D},\Sigma\neq0}
+\frac{1}{4}\mathcal{K}_{14}^{\mathrm{1D},\Sigma\neq0})g_1\tan\theta'
Z^{\prime}_{\mu}\nonumber\\
&&\hspace{-0.2cm}\times\mathrm{tr}[\overline{W}^{\mu\nu}(X_{\nu}\tau^{3}-\tau^{3}X_{\nu})]
+\frac{3i}{4}(\mathcal{K}_{13}^{\mathrm{1D},\Sigma\neq0}
-\frac{1}{4}\mathcal{K}_{14}^{\mathrm{1D},\Sigma\neq0})g_1\tan\theta'
Z^{\prime}_{\mu\nu}tr[X^{\mu}X^\nu\tau^3]\bigg\}\;,\\
&&\hspace{-0.5cm}\mathcal{K}=-\frac{1}{48\pi^2}\left(\log\frac{\kappa^2}{\Lambda^2}+\gamma\right)\hspace{1cm}\Lambda,\kappa\mbox{:
ultraviolet and infrared cutoffs}\;.\label{kappaDef}
\end{eqnarray}


\end{document}